\documentclass[useAMS,usenatbib]{mn2e}

\usepackage{graphicx}
\usepackage[draft]{hyperref} 
\usepackage{enumerate}
\usepackage{xcolor}
\usepackage{longtable}
\usepackage{txfonts}
\usepackage{times}

\newcommand{\hi}{{\sc H\,i}}
\newcommand{\mhi}{$M$(\hi)}

\newcommand{\mJybeam}{mJy beam$^{-1}$}
\newcommand{\Jybeam}{Jy beam$^{-1}$}
\newcommand{\msun}{{M$_\odot$}}
\newcommand{\kms}{$\,$km$\,$s$^{-1}$}
\newcommand{\ltsima} {$\; \buildrel < \over \sim \;$}
\newcommand{\gtsima} {$\; \buildrel > \over \sim \;$}
\newcommand{\lta} {\lower.5ex\hbox{\ltsima}}
\newcommand{\gta} {\lower.5ex\hbox{\gtsima}}

\newcommand{\sofia}{\texttt{SoFiA}}

\title[ASKAP \hi\ imaging of IC~1459]{ASKAP \hi\ imaging of the galaxy group IC~1459}
\author
[P. Serra et al.]{\parbox{\textwidth}{
P. Serra,$^{1}$\thanks{E-mail: \texttt{paolo.serra@csiro.au}}
B. Koribalski,$^{1}$
V. Kilborn,$^{2}$
J. R. Allison,$^{1}$
S. W. Amy,$^{1}$
L. Ball,$^{1}$
K. Bannister,$^{1}$
M. E. Bell,$^{1}$
D. C.-J. Bock,$^{1}$
R. Bolton,$^{1}$
M. Bowen,$^{1}$
B. Boyle,$^{3,1}$
S. Broadhurst,$^{1}$
D. Brodrick,$^{1}$
M. Brothers,$^{1}$
J. D. Bunton,$^{1}$
J. Chapman,$^{1}$
W. Cheng,$^{1}$
A. P. Chippendale,$^{1}$
Y. Chung,$^{1}$
F. Cooray,$^{4,1}$
T. Cornwell,$^{5,1}$
D. DeBoer,$^{6,1}$
P. Diamond,$^{5,1}$
R. Forsyth,$^{1}$
R. Gough,$^{1}$
N. Gupta,$^{7,1}$
G. A. Hampson,$^{1}$
L. Harvey-Smith,$^{1}$
S. Hay,$^{8}$
D. B. Hayman,$^{1}$
I. Heywood,$^{1}$
A. W. Hotan,$^{1}$
S. Hoyle,$^{1}$
B. Humphreys,$^{1}$
B. Indermuehle,$^{1}$
C. Jacka,$^{1}$
C. A. Jackson,$^{9,1}$
S. Jackson,$^{1}$
K. Jeganathan,$^{1}$
S. Johnston,$^{1}$
J Joseph,$^{8}$
P. Kamphuis,$^{1}$
M. Leach,$^{1}$
E. Lenc,$^{10,11,1}$
E. Lensson,$^{1}$
S. Mackay,$^{1}$
M. Marquarding,$^{1}$
J. Marvil,$^{1}$
N. McClure-Griffiths,$^{12,1}$
D. McConnell,$^{1}$
M. Meyer,$^{13}$
P. Mirtschin,$^{1}$
S. Neuhold,$^{1}$
A. Ng,$^{1}$
R. P. Norris,$^{1}$
J. O'Sullivan,$^{1}$
J. Pathikulangara,$^{8}$
S. Pearce,$^{1}$
C. Phillips,$^{1}$
A. Popping,$^{13,1,10}$
R. Y. Qiao,$^{14,8}$
J. E. Reynolds,$^{1}$
P. Roberts,$^{1}$
R. J. Sault,$^{1,15}$
A. E. T. Schinckel,$^{1}$
R. Shaw,$^{1}$
T. W. Shimwell,$^{16,1}$
L. Staveley-Smith,$^{13,10}$
M. Storey,$^{1}$
A. W. Sweetnam,$^{17,1}$
E. Troup,$^{1}$
A. Tzioumis,$^{1}$
M. A. Voronkov,$^{1}$
T. Westmeier,$^{13,1}$
M. Whiting,$^{1}$
C. Wilson,$^{1}$
O. I. Wong,$^{13}$
X. Wu,$^{1}$
}
\vspace{0.4cm}\\ 
\parbox{\textwidth}{
$^{1}$ CSIRO Astronomy and Space Science, Australia Telescope National Facility, PO Box 76, Epping NSW 1710, Australia \\
$^{2}$ Centre for Astrophysics \& Supercomputing, Swinburne University of Technology, PO Box 218, Hawthorn, VIC\\
3122, Australia \\
$^{3}$ SKA Director, Australian Square Kilometre Array Office, Department of Industry and Science, GPO Box 9839,\\
Canberra ACT 2601, Australia \\
$^{4}$  22/1-7 Rowe Street, Eastwood NSW 2122, Australia \\
$^{5}$ SKA Organisation, Jodrell Bank Observatory, Lower Withington, Macclesfield Cheshire SK11 9DL, United Kingdom \\
$^{6}$ Radio Astronomy Laboratory, University of California Berkeley, 501 Campbell, Berkeley CA 94720-3411, USA \\
$^{7}$ Inter-University Centre for Astronomy and Astrophysics, Post Bag 4, Ganeshkhind, Pune University Campus, Pune\\
411 007, India \\
$^{8}$ CSIRO Digital Productivity, PO Box 76, Epping NSW 1710, Australia \\
$^{9}$ International Centre for Radio Astronomy Research (ICRAR), Curtin University, GPO Box U1987, Perth WA 6845,\\
Australia \\
$^{10}$ ARC Centre of Excellence for All-sky Astrophysics (CAASTRO) \\
$^{11}$ Sydney Institute for Astronomy, School of Physics, University of Sydney NSW 2006, Australia \\
$^{12}$ Research School of Astronomy and Astrophysics, Australian National University, Mount Stromlo Observatory,\\
Cotter Road, Weston Creek ACT 2611, Australia \\
$^{13}$ International Centre for Radio Astronomy Research (ICRAR), University of Western Australia, 35 Stirling\\
Highway, Crawley WA 6009, Australia \\
$^{14}$ Sonartech ATLAS Pty Ltd, Unit G01, 16 Giffnock Avenue, Macquarie Park NSW 2113, Australia \\
$^{15}$ School of Physics, University of Melbourne, VIC, 3010, Australia \\
$^{16}$ Leiden Observatory, Leiden University, PO Box 9513, NL-2300 RA Leiden, The Netherlands \\
$^{17}$ 31 Ellalong Road, North Turramurra NSW 2074, Australia \\
}}

\begin{document}

\date{Accepted 2015 June 10.  Received 2015 June 10; in original form 2015 May 12}

\pagerange{\pageref{firstpage}--\pageref{lastpage}} \pubyear{2014}

\maketitle

\label{firstpage}

\clearpage

\begin{abstract}

We present \hi\ imaging of the galaxy group IC~1459 carried out with six antennas of the Australian SKA Pathfinder equipped with phased-array feeds. We detect and resolve \hi\ in eleven galaxies down to a column density of $\sim10^{20}$ cm$^{-2}$ inside a $\sim6$ deg$^2$ field and with a resolution of $\sim1$ arcmin on the sky and $\sim8$ \kms\ in velocity. We present \hi\ images, velocity fields and integrated spectra of all detections, and highlight the discovery of three \hi\ clouds -- two in the proximity of the galaxy IC~5270 and one close to NGC~7418. Each cloud has an \hi\ mass of $\sim10^9$ \msun\ and accounts for $\sim15$ percent of the \hi\ associated with its host galaxy. Available images at ultraviolet, optical and infrared wavelengths do not reveal any clear stellar counterpart of any of the clouds, suggesting that they are not gas-rich dwarf neighbours of IC~5270 and NGC~7418. Using Parkes data we find evidence of additional extended, low-column-density \hi\ emission around IC~5270, indicating that the clouds are the tip of the iceberg of a larger system of gas surrounding this galaxy. This result adds to the body of evidence on the presence of intra-group gas within the IC~1459 group. Altogether, the \hi\ found outside galaxies in this group amounts to several times $10^9$ \msun, at least 10 percent of the \hi\ contained inside galaxies. This suggests a substantial flow of gas in and out of galaxies during the several billion years of the group's evolution.

\end{abstract}

\begin{keywords}
galaxies: evolution, galaxies: ISM.
\end{keywords}

\section{Introduction}
\label{sec:intro}

Galaxy evolution is to a large extent the tale of how galaxies get and lose gas, and how efficiently they convert it into stars. A key part of this tale is the interaction between galaxies and their environment. This can occur through a number of processes such as the continuous streaming of gas from the inter-galactic medium predicted by simulations \citep[e.g.,][]{binney1977,keres2005}, the tidal interaction between galaxies \citep[e.g.,][]{toomre1972} as well as between galaxies and the large-scale gravitational potential \citep[e.g.,][]{bekki2005a}, and the hydrodynamical interaction between galaxies' inter-stellar medium and the inter-galactic medium \citep[e.g.,][]{gunn1972,cowie1977,nulsen1982}. These processes may be largely responsible for determining the rate of gas accretion and removal in galaxies, and for the shifting balance between the populations of early- and late-type galaxies across cosmic time \citep[e.g.,][]{butcher1984} and as a function of environment density \citep[e.g.,][]{dressler1980}.

One of the most direct ways to study these phenomena in the nearby Universe is the observation of galaxies' neutral hydrogen atomic gas (\hi). This gas is often distributed out to a large radius where, if imaged with sufficiently high angular resolution, it can reveal episodes of gas accretion and stripping \citep[e.g.,][]{oosterloo2005,sancisi2008,serra2013}. The resolved observation of \hi\ in galaxies in a wide range of environments --  from large scale voids to galaxy groups and clusters -- has indeed been widely used to study galaxy evolution \citep[e.g.,][]{bravoalfaro2000,verdesmontenegro2001,verheijen2001,chung2009,kreckel2011,serra2012a}. In the future, this type of investigation will be possible on unprecedentedly large areas of the sky, as wide-field \hi\ surveys at  sub-arcminute resolution are planned to be carried out using new radio telescopes such as the Australian Square Kilometre Array Pathfinder (ASKAP; \citealt{johnston2007,johnston2008}) and APERTIF \citep{verheijen2008}.

\begin{table*}
{\centering
\caption{Brightest galaxies in the IC~1459 field, and their properties}
\begin{tabular}{lrrrrr}
\hline
galaxy & $v_\mathrm{hel}$ & $m_B$ & \mhi$_\mathrm{ASKAP}$ & \mhi$_\mathrm{K09}$ & \mhi$_\mathrm{HIPASS}$ \\
  & (km s$^{-1}$) & (mag) & ($10^9$ \msun) & ($10^9$ \msun) & ($10^9$ \msun) \\
(1) & (2) & (3) & (4) & (5) & (6)\\
 \hline
DUKST~406-83 & 1624 &17.1 & - & $0.3\pm0.1$ & - \\
ESO~406-G31      & 1593 & 15.0 & - & - & - \\
ESO~406-G40      & 1238 & 16.3 & $0.5\pm0.1$ & $0.7\pm0.1$ & $1.0\pm0.2$ \\
ESO~406-G42      & 1365 & 15.4 & $3.1\pm0.6$ & $2.2\pm0.2$ & $2.2\pm0.3$ \\
IC~1459$^\mathrm{a}$     & 1802 & 11.0 & - & - & - \\
IC~5264$^\mathrm{a}$     & 1940 & 13.7 & $0.6\pm0.2$ & $1.1\pm0.1$ & - \\
IC~5269            & 1967 & 13.2 & - & - & - \\
IC~5269A$^\mathrm{b}$            & 2870 & 14.2 & $1.4\pm0.3$ & - & $1.4\pm0.3$ \\
IC~5269B$^\mathrm{c}$           & 1667 & 13.2 & $5.1\pm1.0$ & $3.4\pm0.2$ & $5.4\pm0.6$ \\
IC~5269C            & 1783 & 14.2 & $1.1\pm0.3$ & $1.7\pm0.2$ & $1.4\pm0.4$ \\
IC~5270$^\mathrm{c}$              & 1983 & 13.0 & $7.9\pm3.0$ & $7.6\pm0.4$ & $11.4\pm0.5$ \\
IC~5273              & 1293 & 12.2 & $5.4\pm1.1$ & $4.3\pm0.3$ & $4.4\pm0.5$ \\
NGC~7418$^\mathrm{c}$         & 1450 & 12.3 & $4.4\pm0.9$ & $5.5\pm0.3$ & $6.0\pm0.6$ \\
NGC~7418A$^\mathrm{d}$       & 2102 & 13.8 & $4.6\pm0.9$ & $4.7\pm0.3$ & $4.7\pm0.4$ \\
NGC~7421         & 1792 & 13.0 & $1.1\pm0.3$ & $0.9\pm0.1$ & $1.0\pm0.3$ \\
2dFGRS~S537Z045 & 2188 & 16.2 & - & - & - \\
2MASX~J22571092-3640103$^\mathrm{a,d}$ & 1945 & 15.8 & - & - & - \\
\hline
\label{tab:memb}
\end{tabular}
}

\it Column 1. \rm Galaxy name. \it Column 2. \rm Heliocentric velocity from the NASA Extragalactic Database (NED). \it Column 3. \rm $B$-band apparent magnitude from NED. \it Column 4. \rm \hi\ mass from this work. This is calculated assuming a distance of 29 Mpc for all galaxies (see text). \it Column 5. \rm \hi\ mass from \cite{kilborn2009} scaled to the distance of 29 Mpc assumed here. \it Column 6. \rm \hi\ mass re-measured by us using the HIPASS data \citep{barnes2001}. \it Notes \rm (a) \cite{kilborn2009} associate the \hi\ detection GEMS\_IC1459\_9 with the three galaxies IC~1459, IC~5264 and 2MASX~J22571092-3640103. Higher resolution \hi\ images by \cite{walsh1990} and \cite{oosterloo1999} show that part of this gas belongs to IC~5264 while the rest is distributed in the intra-group medium. (b) This galaxy is outside the velocity range covered by \cite{kilborn2009} and therefore no \mhi\ comparison is possible. (c) The source is spatially resolved in the HIPASS data, and the \hi\ mass value is estimated by fitting a 2D elliptical Gaussian to the object. (d) \cite{kilborn2009} associate the \hi\ detection GEMS\_IC1459\_7 with the two galaxies NGC~7418A and 2MASX~J22571092-3640103. Based on higher resolution \hi\ images by \cite{walsh1990} and \cite{oosterloo1999} we associate all the \hi\ with NGC~7418A.
\end{table*}

The WALLABY survey will make use of the wide-field capabilities of ASKAP to image \hi\ in galaxies at $z<0.25$  across 3/4 of the sky in just one year of observing time \citep{koribalski2012b}. The key advantage of ASKAP over traditional radio interferometers is that it is equipped with phased-array feeds \citep[e.g.,][]{hay2008}. These allow observers to simultaneously form multiple beams on the sky and, therefore, observe a much larger field of view than possible with feed horns. Leading up to WALLABY we have carried out a number of commissioning \hi\ observations with the ASKAP prototype -- the Boolardy Engineering Test Array. This consists of 6 ASKAP antennas equipped with first-generation phased-array feeds ($T_\mathrm{sys}/\eta \sim180$ K at 1.4 GHz), and allows us to form 9 simultaneous dual-polarisation beams on the sky. We refer to \cite{hotan2014} for details on this system.

In this paper we study the \hi\ content of the galaxy group IC~1459 using ASKAP data. We detect \hi\ in eleven galaxies and are able to resolve their \hi\ morphology and kinematics down to a column density of $\sim10^{20}$ cm$^{-2}$ over a $\sim6$ deg$^2$ field of view, with a resolution of $\sim1$ arcmin on the sky and $\sim8$ \kms\ in velocity. Below we summarise the main properties of IC~1459 (Sec. \ref{sec:ic1459}), describe the radio observations and data reduction (Sec. \ref{sec:data}), discuss the \hi\ content of galaxies in the group (Sec. \ref{sec:results}) and summarise our findings (Sec. \ref{sec:summary}).

\begin{figure*}
\includegraphics[width=18cm]{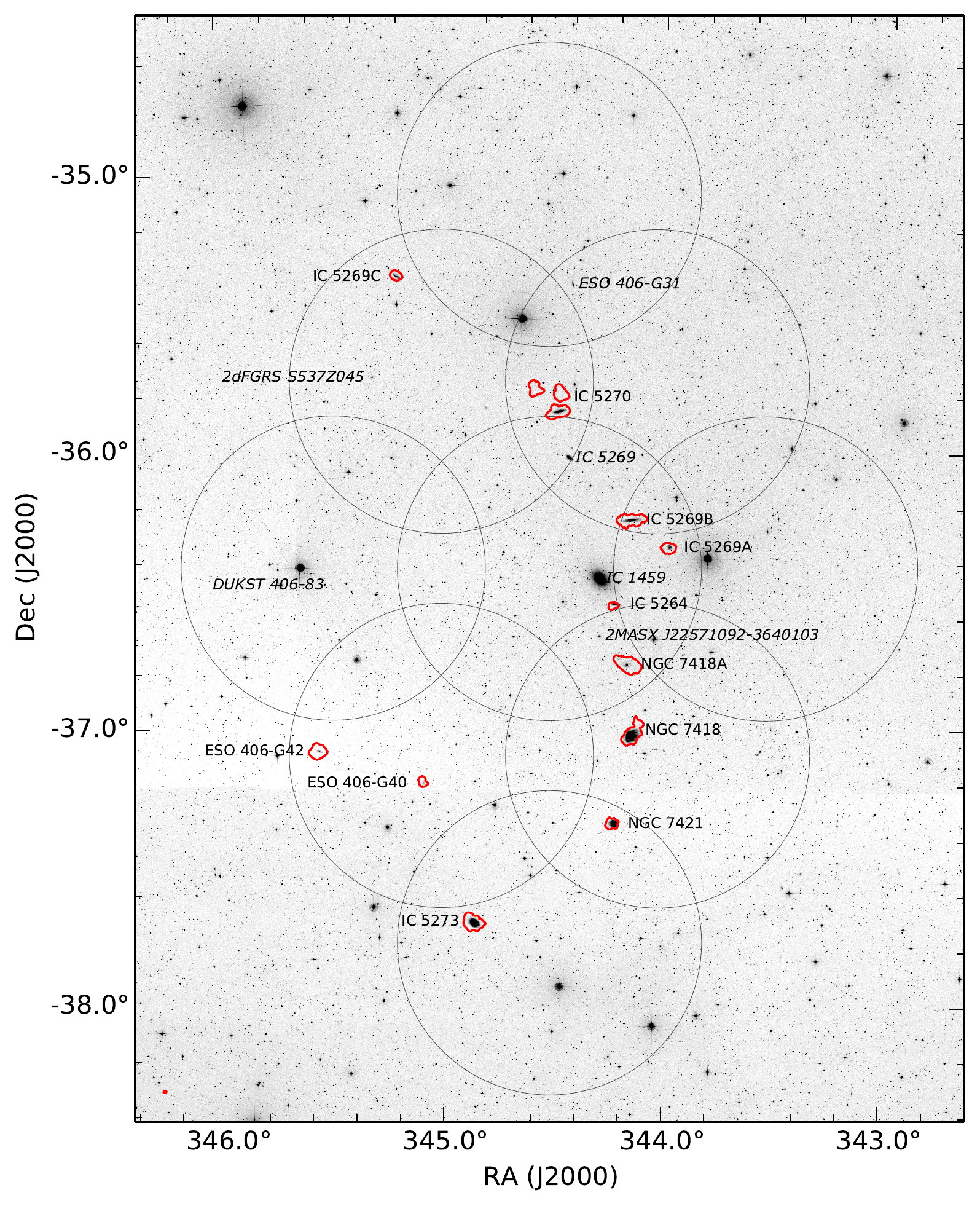}
\caption{ASKAP \hi\ contours (in red) overlaid on the DSS2-red image of the IC~1459 group. The \hi\ contour level corresponds to a column density of $10^{20}$ cm$^{-2}$, equivalent to 0.35 \Jybeam \kms. The PSF of the \hi\ image is represented by the red ellipse in the bottom-left corner. The grey circles indicate the position of the 9 beams. They have a diameter of $1.1^\circ$, equal to the beams' FWHM at 1.4 GHz. We label galaxies listed in Table \ref{tab:memb} using normal and italic fonts for \hi\ detections and non-detections, respectively. Labels are located to the east (west) of galaxies on the east (west) side of the field.}
\label{fig:mosaic}
\end{figure*}

\section{The galaxy group IC~1459}
\label{sec:ic1459}

IC~1459 is a loose galaxy group hosting ten bright galaxies of mostly late morphological type \citep{brough2006}. It is named after its central, luminous early-type member, which has a distance of 29 Mpc \citep{blakeslee2001,tonry2001}. The group is part of the larger-scale Grus cluster -- a loose over-density hosting mostly spirals \citep{aaronson1981}. A number of additional, fainter galaxies might be associated with IC~1459 based on their recessional velocity \citep[e.g.,][]{kilborn2009}. Table \ref{tab:memb} lists all 17 galaxies brighter than $m_B\sim17$ mag and with recessional velocity below 4000 \kms\ within the field of view of our observations. Fig. \ref{fig:mosaic} shows an image of the group. With the exception of IC~5269A, all galaxies are within $\pm3\times\sigma_\mathrm{group}$ from the central early-type IC~1459, where $\sigma_\mathrm{group}=220$ \kms\ is the group velocity dispersion \citep{brough2006}. We adopt the same distance of 29 Mpc for all galaxies in the Table but note that IC~5269A may be in the background (its distance is 37 Mpc according to \citealt{springob2009}).

The diffuse intra-group medium of IC~1459 is detected in X-rays \citep{osmond2004}. This dense medium should create an environment hostile to the survival of \hi\ in the group. Yet IC~1459 is dominated by gas-rich, blue spirals. This suggests that the group may be in a relatively early stage of its assembly. Along the same line, there appear to be kinematically distinct subgroups as the 5 southernmost galaxies in Table \ref{tab:memb} all have recessional velocity below that of the central early-type (ESO~406-G40, ESO~406-G42, IC~5273, NGC~7418, NGC~7421). The observation of \hi\ in the group could help clarify to what extent its members are interacting with one another as well as with the group's gravitational potential and gaseous medium.

\cite{kilborn2009} analyse the \hi\ content of IC~1459 using the Parkes telescope as part of their study of galaxy groups in the GEMS sample \citep{osmond2004}. They report 18 \hi\ detections over a $5.5^\circ\times5.5^\circ$ field -- all unresolved by the $\sim15$ arcmin Parkes beam and all with an optical counterpart. They also find that galaxies in the group have fairly typical \hi\ mass for their morphology, suggesting that gas removal in the group is not significant. In contrast with this result, an earlier study by \cite{sengupta2006} based on shallower data from the \hi\ Parkes All Sky Survey \citep{barnes2001} finds that the group members IC~5264 (a dusty early type), IC~5269B, IC~5269C, NGC~7418 and NGC~7421 (all late types) are \hi\ deficient. This disagreement is most likely due to the large uncertainty on the \hi\ deficiency of individual galaxies. More precise indications about the occurrence of gas removal in IC~1459 may come from the resolved study of the \hi\ morphology with interferometry. This has been done for a few galaxies in the group.

\cite{ryder1997} discuss signatures of ram pressure stripping in NGC~7421 using an \hi\ image obtained with the Australia Telescope Compact Array (ATCA). They also mention (but do not show) the warped appearance of the \hi\ disc of NGC~7418. Interestingly, data taken with both the Very Large Array \citep{walsh1990} and the ATCA \citep{oosterloo1999} reveal the presence of low-column-density ($<10^{20}$ cm$^{-2}$), intra-group \hi\ near the central galaxy IC~1459. This gas may have been lost by one of the spirals as it travelled across the group and could be related to the disturbed appearance of the early-type galaxy itself -- it exhibits twisted isophotes \citep{williams1979}, a counter-rotating stellar core \citep{franx1989} and low-surface-brightness spiral-like features in deep optical imaging \citep[][see \url{http://ftp.aao.gov.au/images/deep_html/i1459_gr_d.html}]{malin1985}. This galaxy also hosts an active galactic nucleus charactertised by two symmetric radio jets ($\sim1$ Jy at 1.4 GHz; \citealt{tingay2015}), whose activity may have been triggered by the same events that gave the galaxy its peculiar morphology and kinematics. This body of evidence indicates that at least some interaction between galaxies is occurring within the group.

Finally, and although not the main focus of their respective articles, both \cite{walsh1990} and \cite{oosterloo1999} show \hi\ images of IC~5264, IC~5269B and NGC~7418A. The image of \cite{oosterloo1999} also includes IC~5269A. IC~5269B is imaged in \hi\ by \cite{sengupta2007}, too, using data taken with the Giant Metrewave Radio Telescope. Among these systems, only NGC~7418A shows a disturbed gas morphology.

The field of view of the \hi\ images mentioned above is too small to include other galaxies in the group. To the best of our knowledge, this paper is the first to show the \hi\ morphology of ESO~406-G40, ESO~406-G42, IC~5269C, IC~5270, IC~5273 and NGC~7418.

\section{Observations and data reduction}
\label{sec:data}

\subsection{Beam forming and observations}
\label{sec:beamform}

We observe IC~1459 for a total of 30 h with ASKAP. This integration is divided into three individual observations of $\sim10$ h carried out on 28 August, 2 September and 10 September 2014, respectively. Observations are performed at night. They cover the frequency range 1.223 to 1.527 GHz. This 304 MHz band is divided into 16,416 contiguous channels of width $\sim18.5$ kHz, corresponding to $\sim3.9$ \kms\ for \hi\ at $z\sim0$. 

We distribute the 9 dual-polarisation beams available with ASKAP as shown in Fig. \ref{fig:mosaic}. This footprint is chosen to include as many known \hi\ systems as possible. The spacing between beams is $0.78^\circ$ ($\sim 70$ percent of the beam FWHM at 1.4 GHz). This is smaller than the $\sim1^\circ$ spacing planned for future surveys, which will tile the $5.5^\circ\times5.5^\circ$ field of view of ASKAP with 36 beams. The adopted spacing delivers a more uniform noise level across the observed field. 

Beams are formed by computing a set of complex weights before the first and the third observation. The second observation uses the same beam weights as the first one. Briefly, for each of the 9 beams we position the Sun at the centre of the beam, observe it for 2 minutes and calculate the complex weights which give the maximum signal-to-noise ratio (S/N) at the beam centre, referenced to a noise measurement with the dish pointing $15^\circ$ south of
 the Sun. This calculation is performed in 64 1-MHz-wide channels distributed over the 304 MHz bandwidth and unevenly separated by an interval of 4 or 5 MHz. We refer to \cite{hotan2014} for more details.

\subsection{Bandpass and gain calibration}
\label{sec:bpgain}

For the purpose of this work we reduce and analyse only 1,000 channels in the frequency range 1.4025 to 1.4210 GHz (18.5 MHz bandwidth). Data within this frequency range are essentially free from radio frequency interference, and flagging with a high amplitude threshold is sufficient to remove the few bad visibilities.

We reduce the data using the \texttt{MIRIAD} package \citep{sault1995}. In order to calibrate the bandpass and set the flux scale we observe PKS~B1934-638 at the centre of each beam before each 10-h observation. We integrate 15 min per beam (135 min total). We calibrate each observation separately and assume that the bandpass does not vary during the 10 hours.

For each 10-h observation and each beam we calibrate the antenna-based gain phases as a function of time in two steps. First, we calibrate with a sky model built from the NRAO VLA Sky Survey (NVSS) image of the field \citep{condon1998}. We construct the sky model by applying the ASKAP beam response at the appropriate position of the NVSS image. In doing so we assume a Gaussian beam with FWHM of $1.1^\circ$ at 1.4 GHz. The sky model consists of all pixels above $\sim10 \ \sigma$ in the resulting image. We treat these pixels as clean components and calibrate the gains on a time interval of 30 minutes. This initial calibration is sufficient to produce a reasonably good continuum image. In the second step of our procedure we clean this image and use the clean components to self-calibrate the gain phases on a time interval of 1 minute.  The calibrated data are imaged and cleaned again, resulting in the final continuum image.

\begin{figure}
\includegraphics[width=8.5cm]{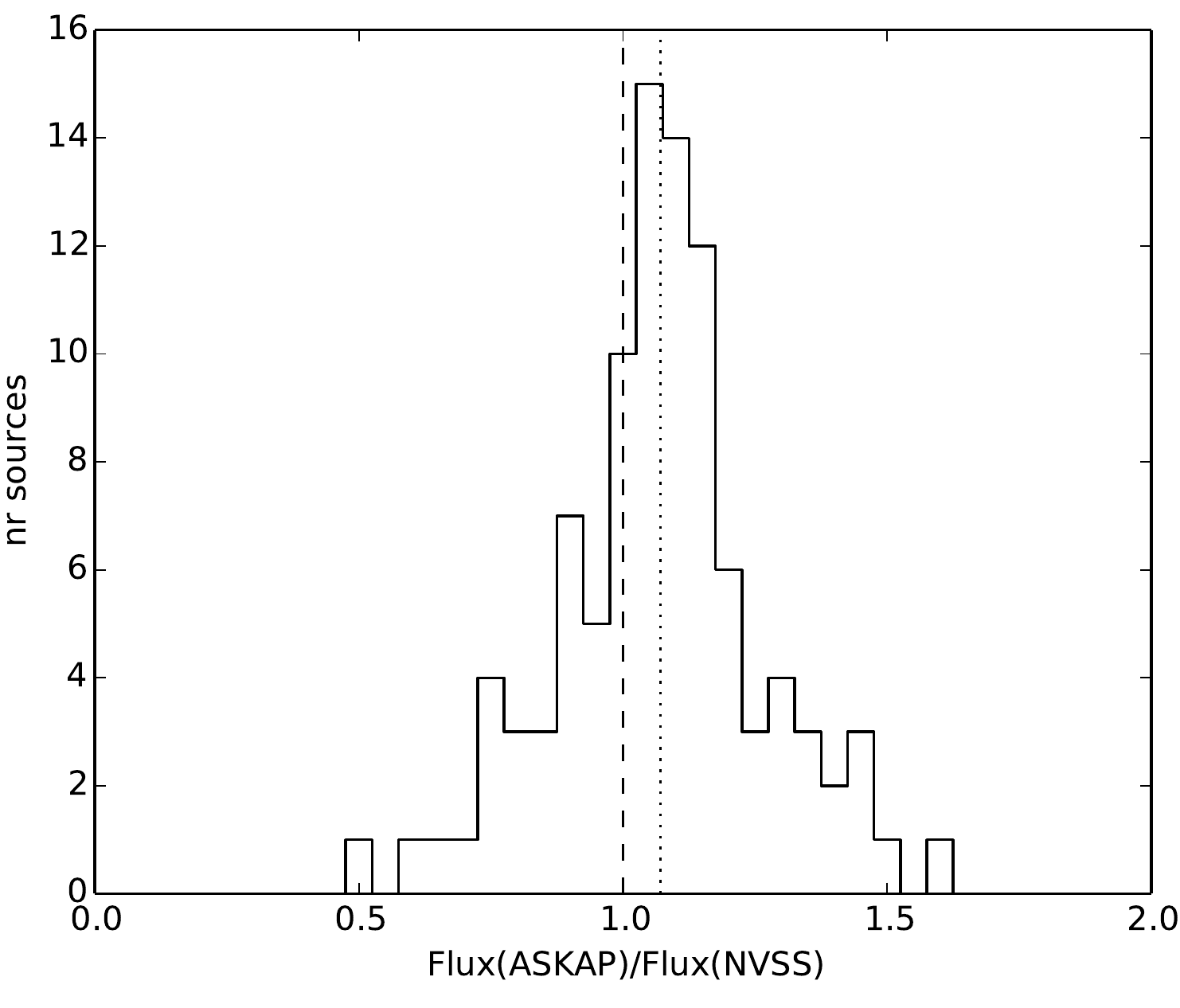}
\caption{Distribution of the ASKAP-to-NVSS peak flux ratio for 100 radio continuum sources detected above $10\sigma$ in the ASKAP continuum mosaic. The ASKAP continuum mosaic is obtained using a 18.5-MHz-wide band between 1.4025 and 1.4210 GHz. The dashed line indicates a ratio of 1. Both mean and median of the distribution are 1.07 (dotted line). The standard deviation is 0.20.}
\label{fig:vast}
\end{figure}

We evaluate the quality of our calibration/imaging procedure by comparing a linear mosaic formed with all single-beam, 10-h continuum images to NVSS. We build the continuum mosaic making the same assumptions about the ASKAP beam shape as above, and giving all images equal weights. (We describe a more accurate mosaicking method in Sec. \ref{sec:hiimaging} when discussing the \hi\ mosaic.) The comparison with NVSS allows us to test the combined effect of our assumptions about the beam shape and size, which we implicitly take to be identical for all 9 beams on all 6 antennas and to be constant with time. The astrometry of our continuum mosaic is tied to that of NVSS by our gain calibration strategy (see above). Therefore, the good positional match between NVSS and ASKAP sources (a few arcsec) is expected. A more interesting comparison is that between the ASKAP and NVSS peak fluxes (Fig. \ref{fig:vast}). We find an average ASKAP-to-NVSS peak flux ratio of 1.07. The standard deviation is $\sim0.2$. This scatter is likely to reflect the limitations of our assumptions about the beams. However, it is comparable to the typical uncertainty on the \hi\ mass of distant galaxies and, therefore, it has no significant impact on this work.

\subsection{\hi\ imaging}
\label{sec:hiimaging}

For each 10-h observation and each beam we obtain a spectral-line dataset by subtracting the clean components of the final continuum image from the calibrated visibilities. We subtract any remaining continuum emission from the visibilities using the \texttt{MIRIAD} task UVLIN. We fit a linear function excluding the central third of the band, which corresponds to a recessional velocity range of $\sim1200$ to $\sim2500$ \kms\ for the \hi\ line.

We image \hi\ in each beam by gridding all three 10-h continuum-subtracted visibility sets onto the same \it uv \rm grid before Fourier transforming. We make \hi\ cubes with a channel width of 8 \kms, and apply natural weighting to obtain the maximum sensitivity to low-column-density \hi\ emission. The large difference in baseline length between the shortest and the second-shortest baseline \citep[37 m vs 144 m;][]{hotan2014} results in a natural point spread function (PSF) with high, broad sidelobes. This makes cleaning difficult, and for this reason we also make natural-weighted \hi\ cubes excluding the shortest baseline. This lowers the number of baselines from 15 to 14, increasing the noise level in the resulting cubes by just a few percent. However, the PSF area is $\sim1.3$ times smaller and, therefore, the column density sensitivity is worse by a similar factor. We base our study of the IC~1459 galaxy group on the latter cubes, although we inspect visually the cubes obtained including the 37-m baseline to look for additional faint, diffuse \hi\ emission (see Sec. \ref{sec:results}).

For each beam we construct a clean mask by selecting bright voxels in the \hi\ cube smoothed with a circular Gaussian filter with FWHM of 1 arcmin and a Hanning filter of width 5 channels. We clean the \hi\ cubes within these masks down to the r.m.s. noise in the 9 individual cubes ($\sim9$ \mJybeam). We finally restore the clean components with an elliptical Gaussian PSF with major- and minor axis FWHM of $\sim70$ arcsec and $\sim55$ arcsec, respectively, and $\mathrm{PA}=-80^\circ$. The formal $5\sigma$ column density sensitivity of these cubes is $\sim1.5\times10^{20}$ cm$^{-2}$ within 16 \kms.

The 9 deconvolved \hi\ cubes are combined to form the final \hi\ mosaic cube. In standard mosaicking, deconvolved cubes (or images) are combined by linearly weighting each pixel by the inverse of the variance \citep[e.g.,][]{cornwell1988}. Under the assumption of independent cubes, this method minimises the noise level in the mosaic. However, cubes created using data from phased-array feeds are not necessarily independent and a more general approach is needed. The mosaic with minimum noise level is given by:

\begin{equation}
I_\mathrm{mosaic}(l,m,\nu)=\frac{\mathbf{B}^\mathrm{T}(l,m,\nu)\, \mathbf{C}^{-1}(\nu)\, \mathbf{I}(l,m,v)}{\mathbf{B}^\mathrm{T}(l,m,\nu)\, \mathbf{C}^{-1}(\nu)\, \mathbf{B}(l,m,\nu)},
\label{eq1}
\end{equation}

\noindent where for each position $(l,m,\nu)$ in the cube, \bf I \rm and \bf B \rm are $N\times1$ matrices representing the $N$ cubes to be mosaicked and the $N$ beams, respectively, and \bf C \rm is the $N\times N$ image-plane noise covariance matrix (in our case $N=9$). This expression is equivalent to the traditional inverse-variance weighting if \bf C \rm is diagonal.

As above, we assume that the beams are Gaussian with $\mathrm{FWHM}=1.1^\circ$ independent of frequency within the 18.5 MHz band under consideration, and truncate them at a response level of 25 percent. Similarly, we assume that \bf C \rm does not depend on frequency. We calculate \bf C \rm separately for 50 channels devoid of \hi\ emission in the individual cubes, and take a median across frequency. Under these assumptions, we obtain the frequency-independent mosaic noise image:

\begin{equation}
\sigma_\mathrm{mosaic}(l,m)=(\mathbf{B}^\mathrm{T}(l,m)\, \mathbf{C}^{-1}\, \mathbf{B}(l,m))^{-1/2}.
\label{eq2}
\end{equation}

\noindent We measure correlation coefficients of 0.13 to 0.20 for pairs of adjacent beams, which are separated by $0.78^\circ$ (see Sec. \ref{sec:beamform}). The variation from pair to pair may indicate that the beams are not circularly symmetric. The correlation is negligible for all other beam pairs. The noise level in the final mosaic \hi\ cube is $\sim 9.5$ \mJybeam\ $\pm10$ percent over a $\sim4$ deg$^2$ field. This is consistent with the $T_\mathrm{sys}/\eta\sim180$ K reported by \cite{hotan2014}. The noise level remains below twice this value within $\sim 6$ deg$^2$ in the mosaic \hi\ cube. Using the more common inverse-variance weighting instead of Eq. \ref{eq1} would result in a mosaic with r.m.s. noise level $\sim0.5$ \mJybeam\ higher in the inner part of the field.

\begin{figure*}
\includegraphics[width=5cm]{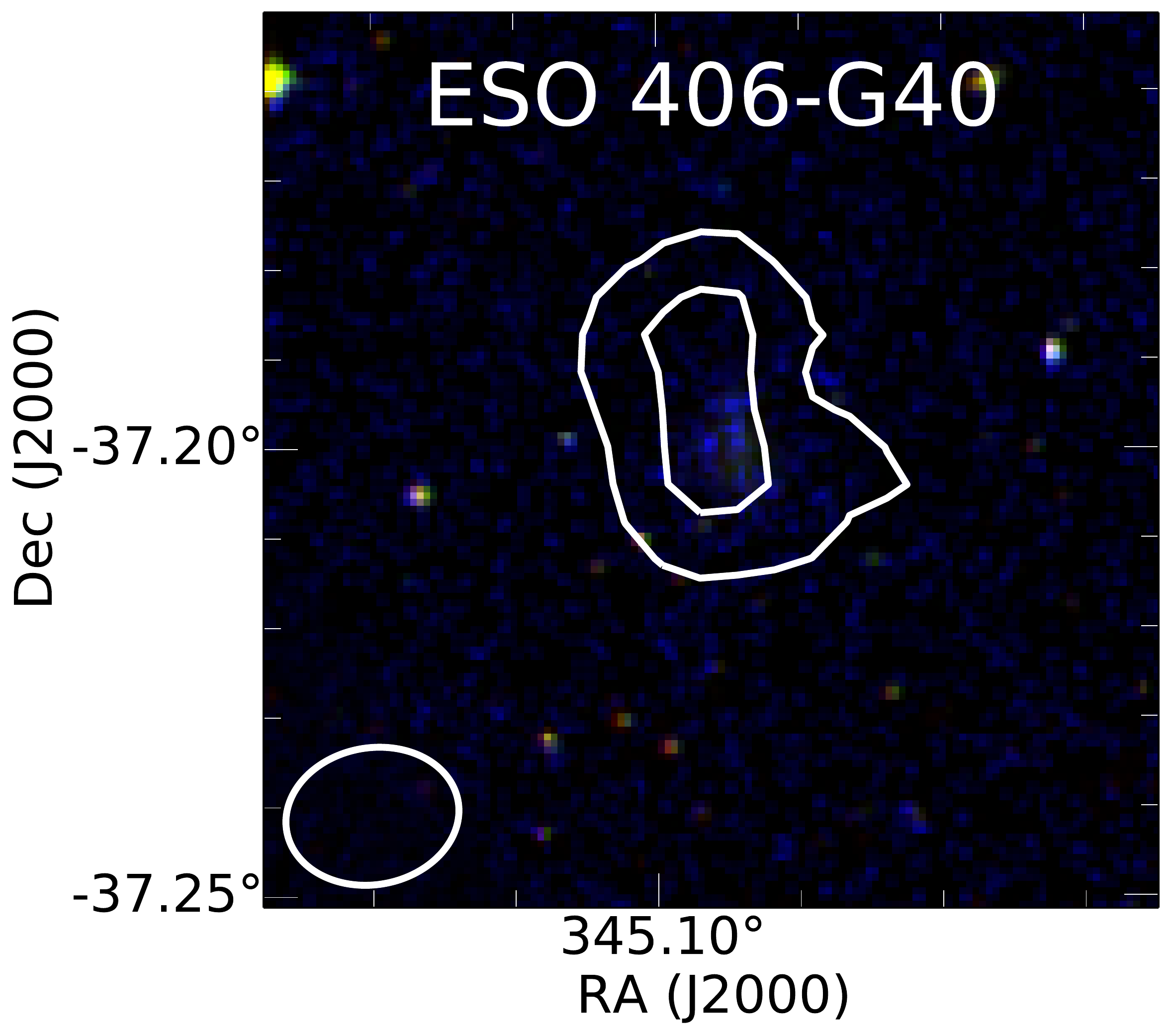}
\includegraphics[width=5cm]{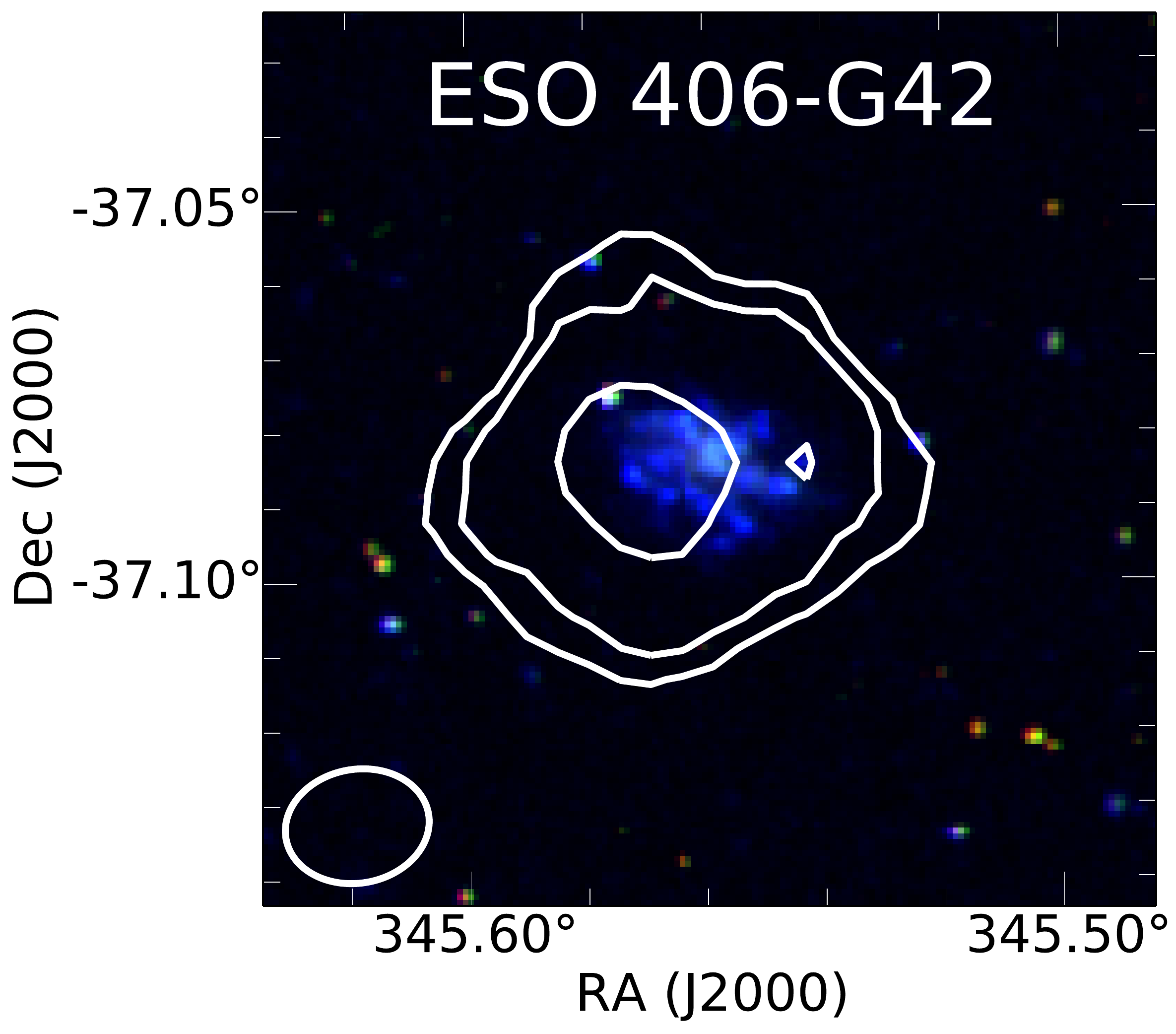}
\includegraphics[width=5cm]{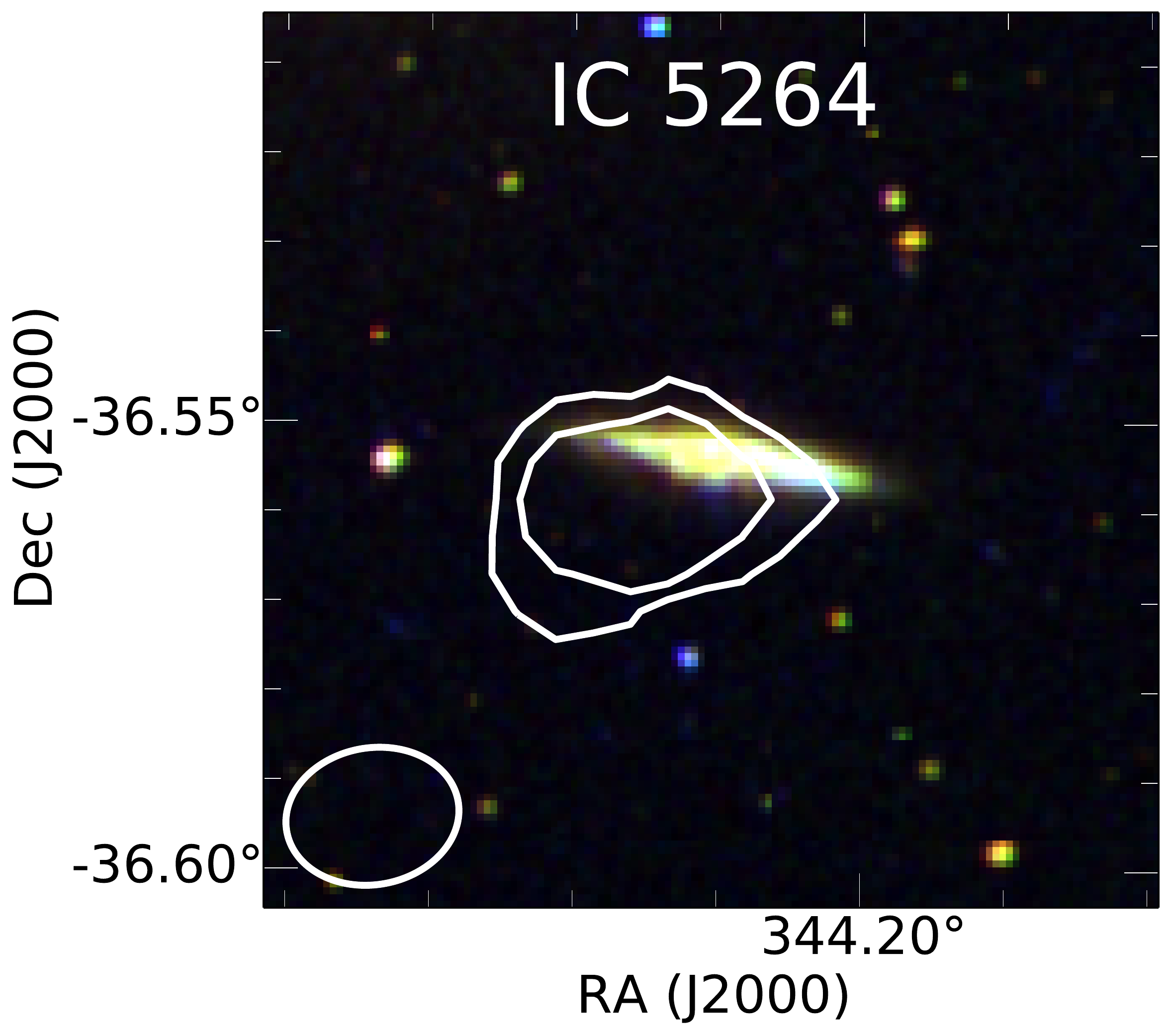}
\includegraphics[width=5cm]{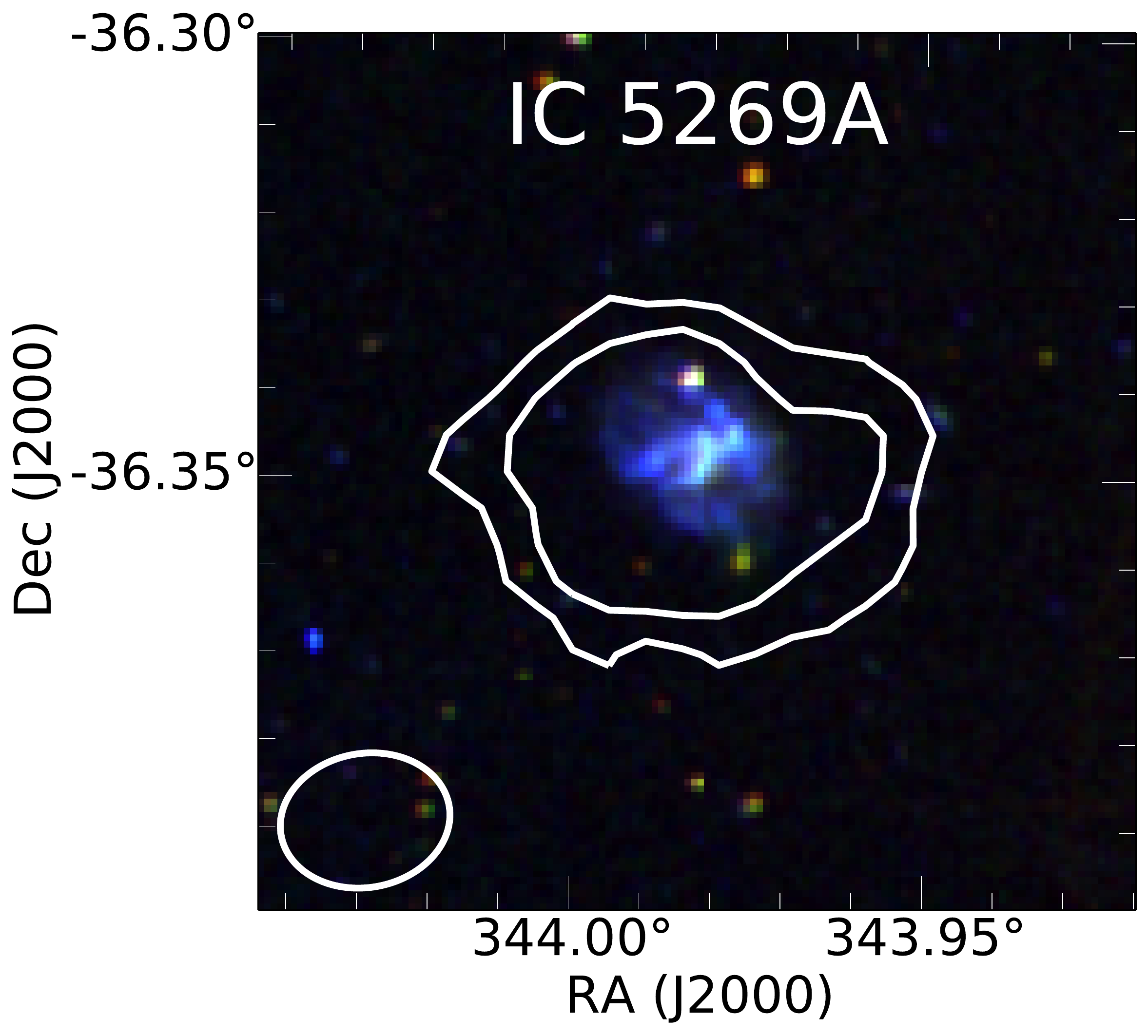}
\includegraphics[width=5cm]{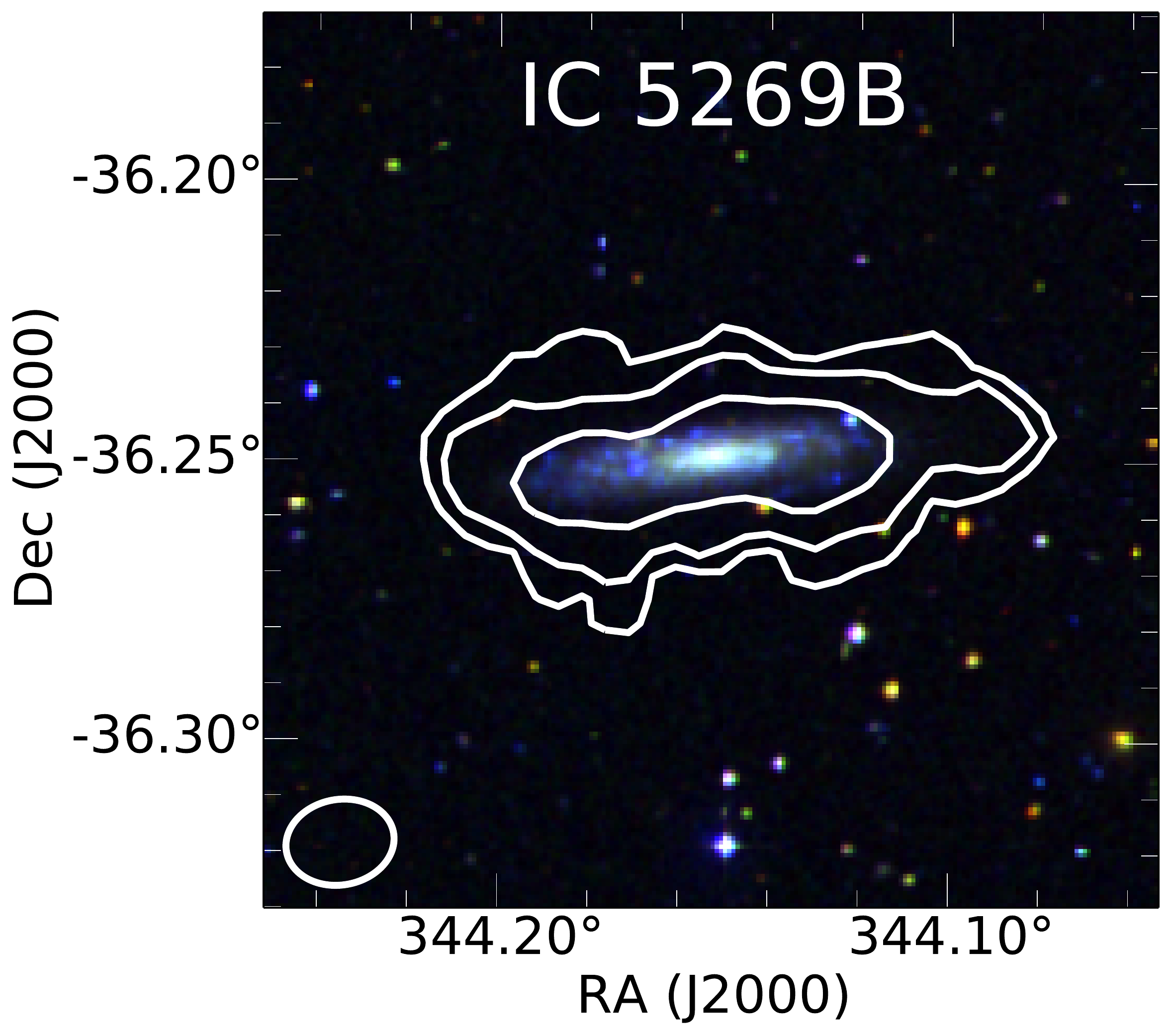}
\includegraphics[width=5cm]{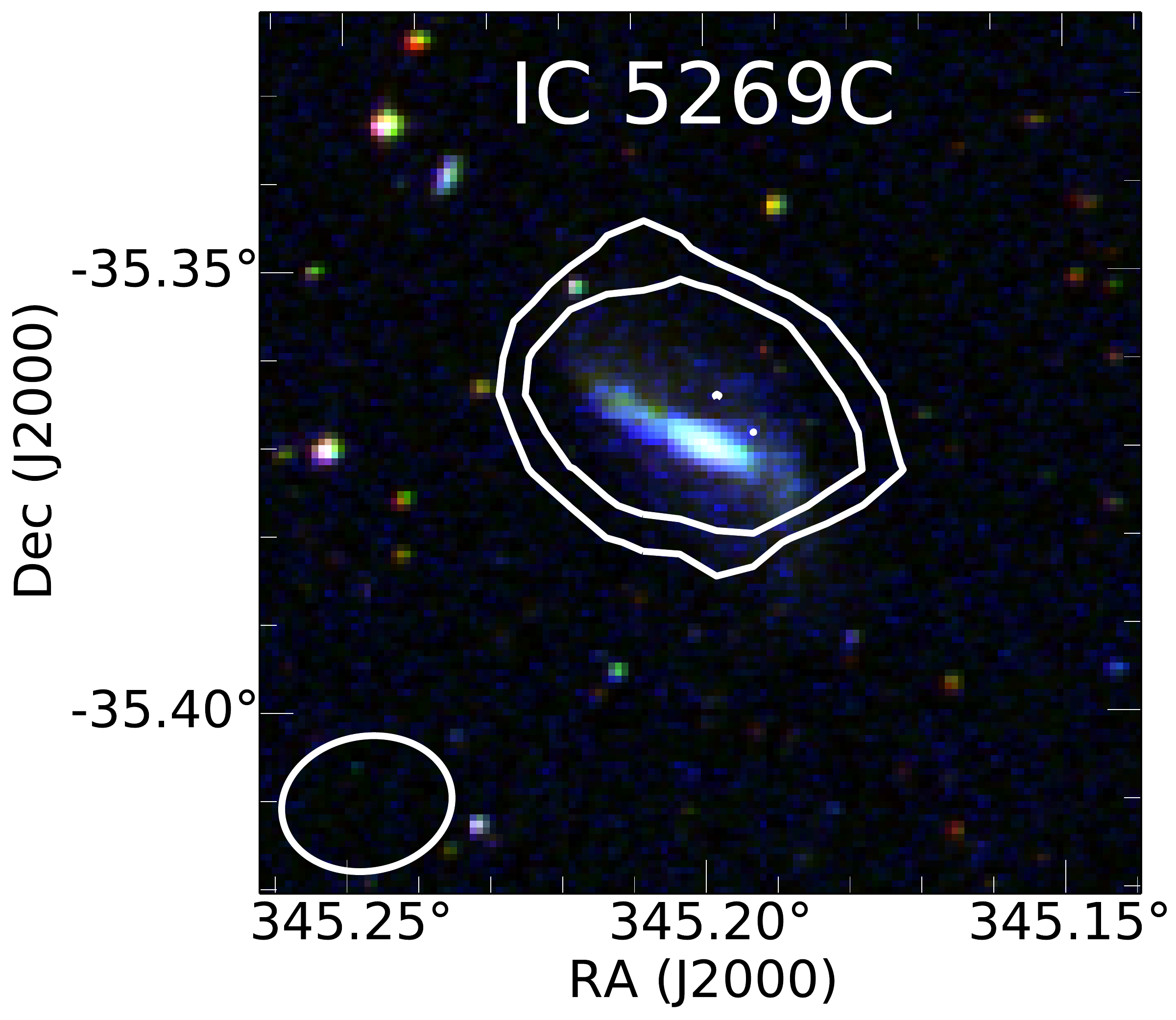}
\includegraphics[width=5cm]{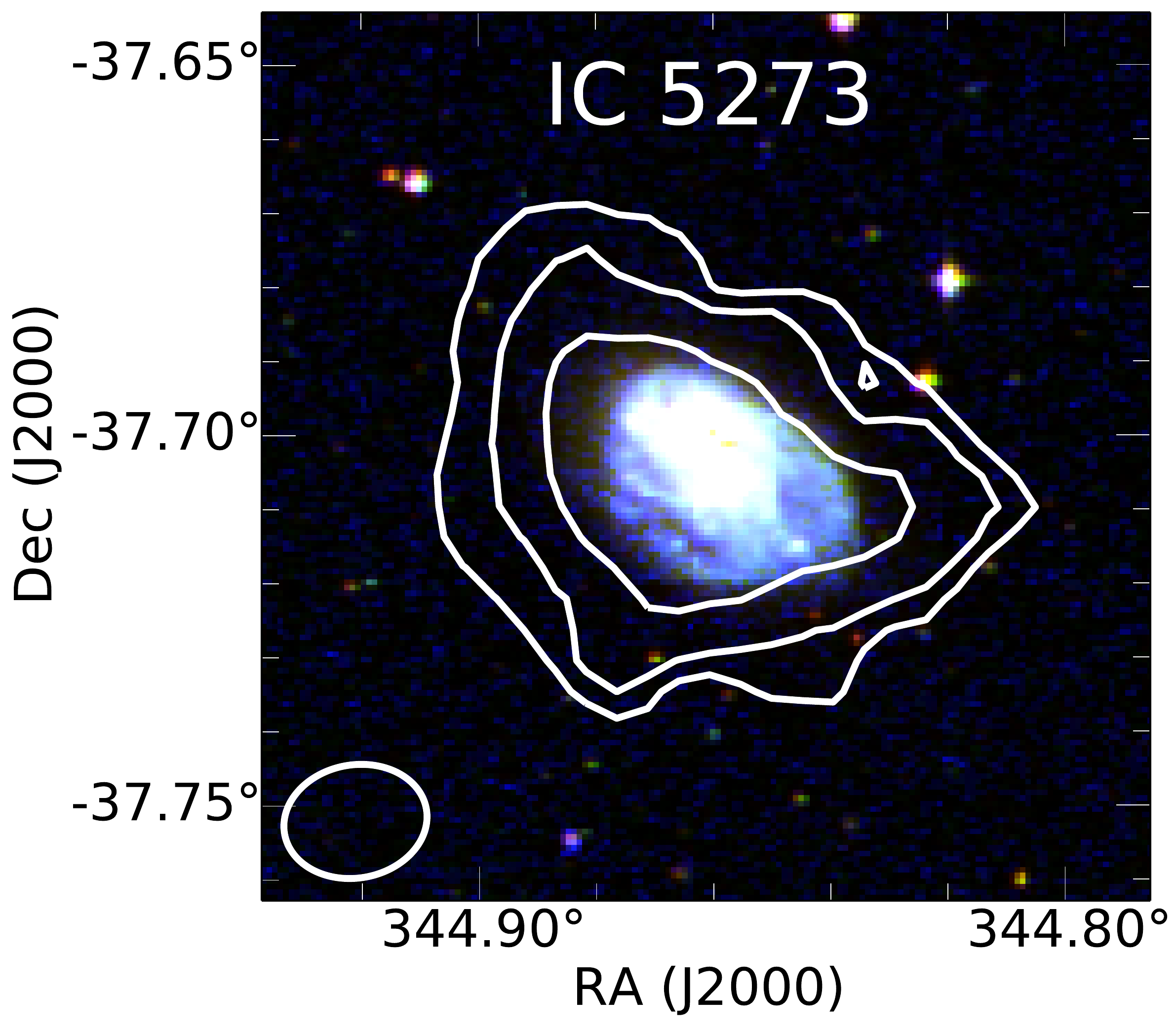}
\includegraphics[width=5cm]{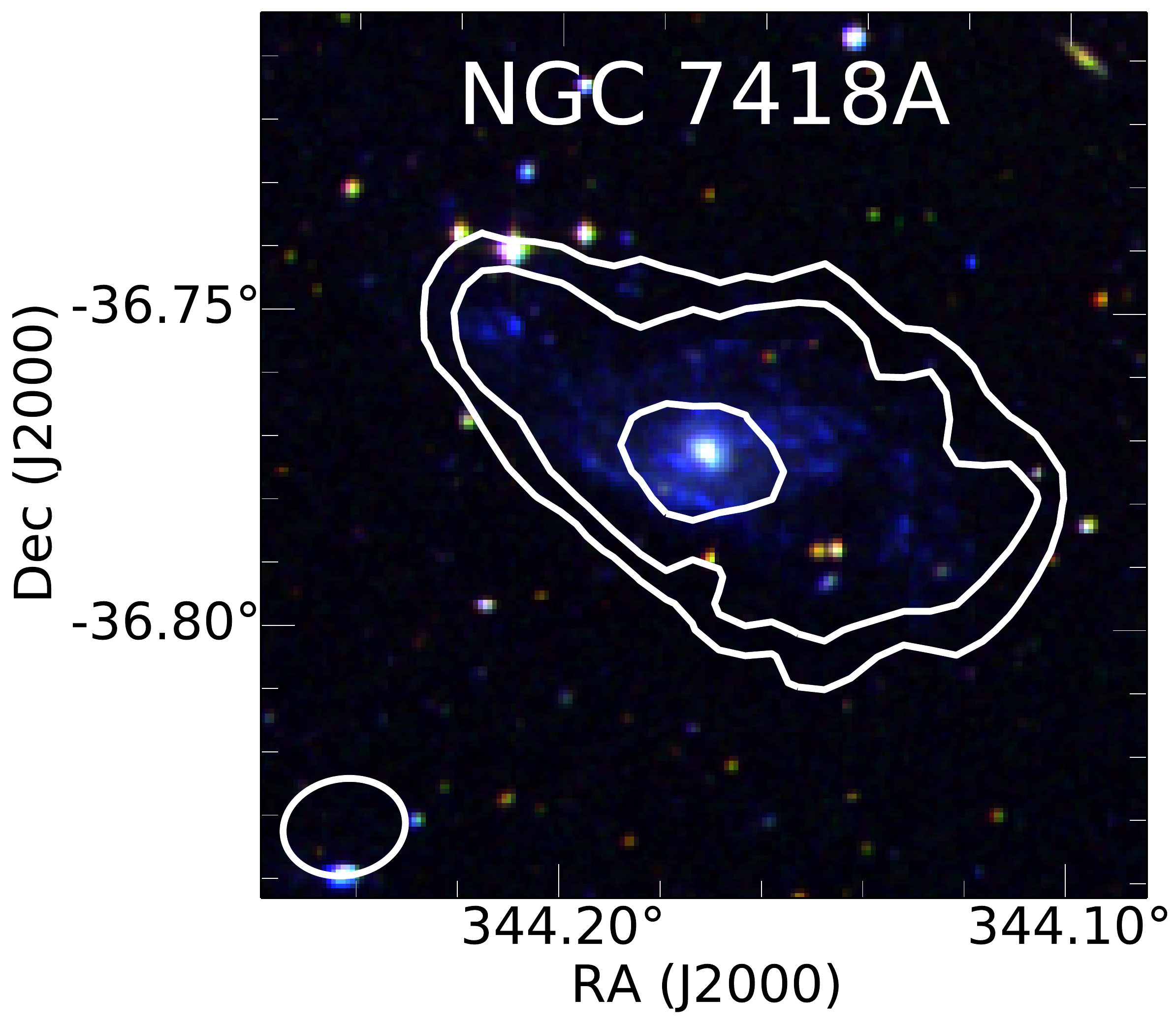}
\includegraphics[width=5cm]{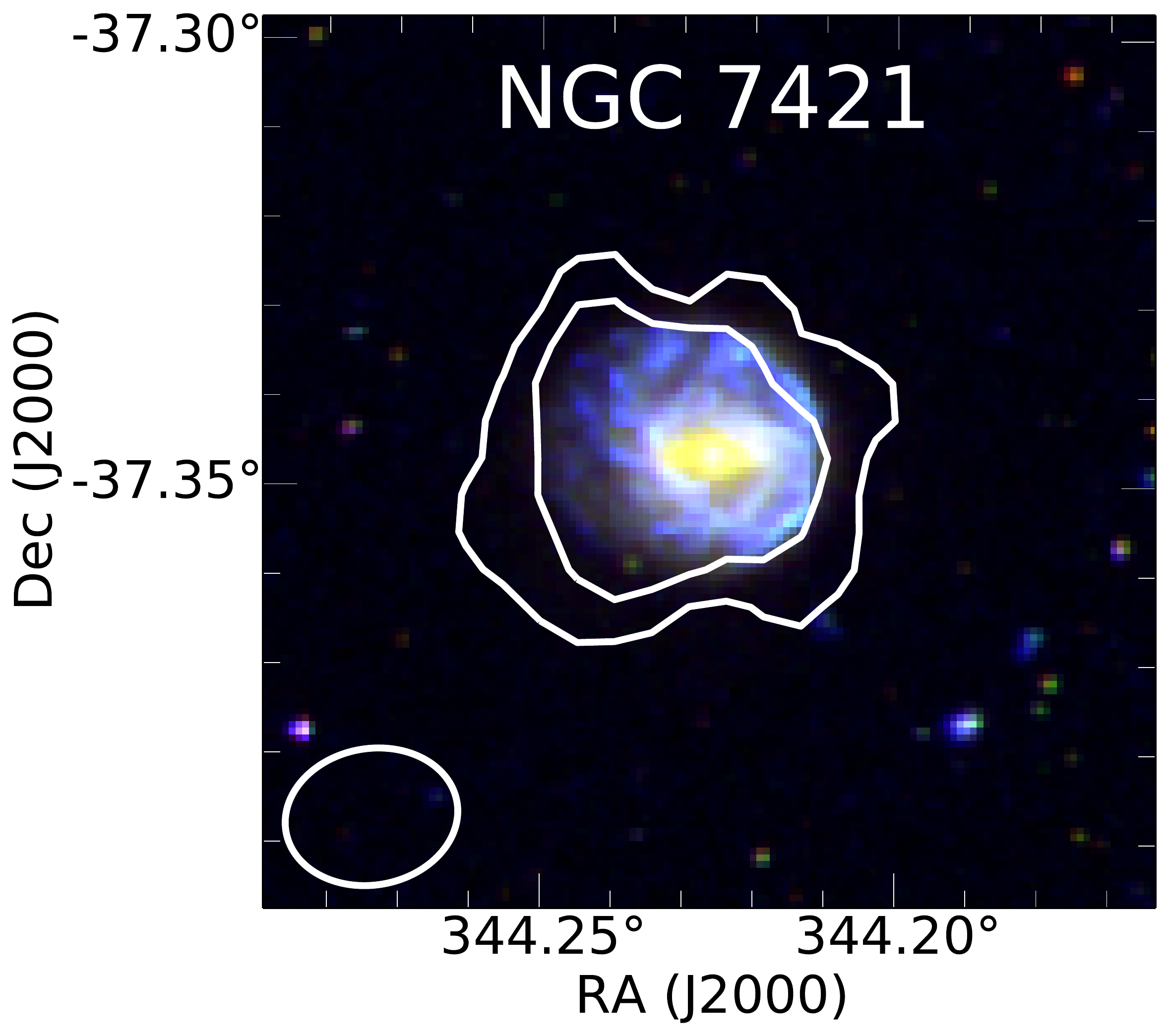}
\caption{ASKAP \hi\ contours (white) overlaid on an RGB image of nine of the eleven detected galaxies (for IC~5270 and NGC~7418 see Fig. \ref{fig:ic5270}). The RGB images are obtained using publicly available DSS2-red (R), DSS2-blue (G) and GALEX NUV (B) images. The contour levels are $3^n\times10^{20}$ cm$^{-2}$ ($n=0,1,2$). The PSF is shown in the bottom-left corner.}
\label{fig:cutouts0}
\end{figure*}

\begin{figure*}
\includegraphics[width=5cm]{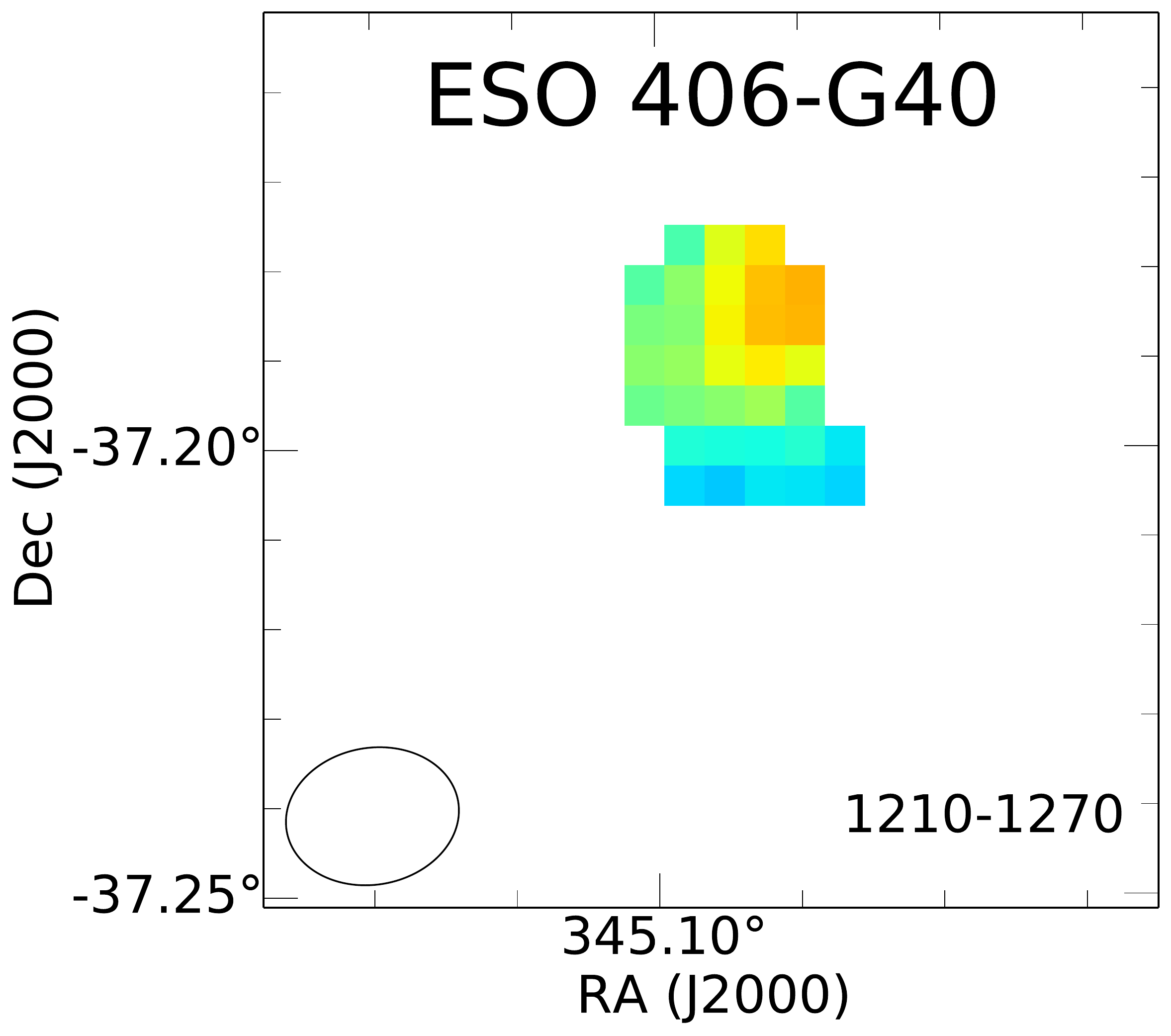}
\includegraphics[width=5cm]{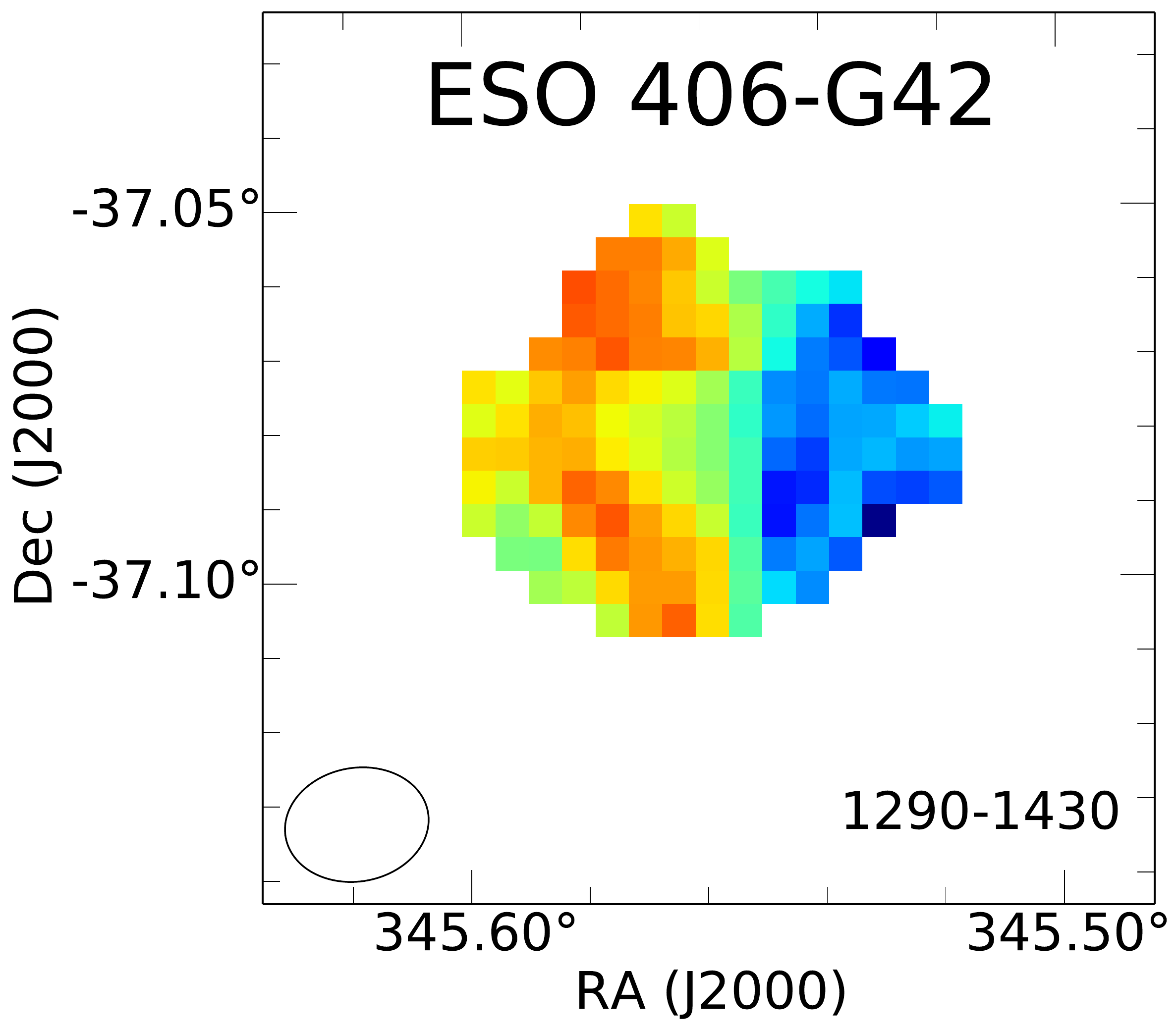}
\includegraphics[width=5cm]{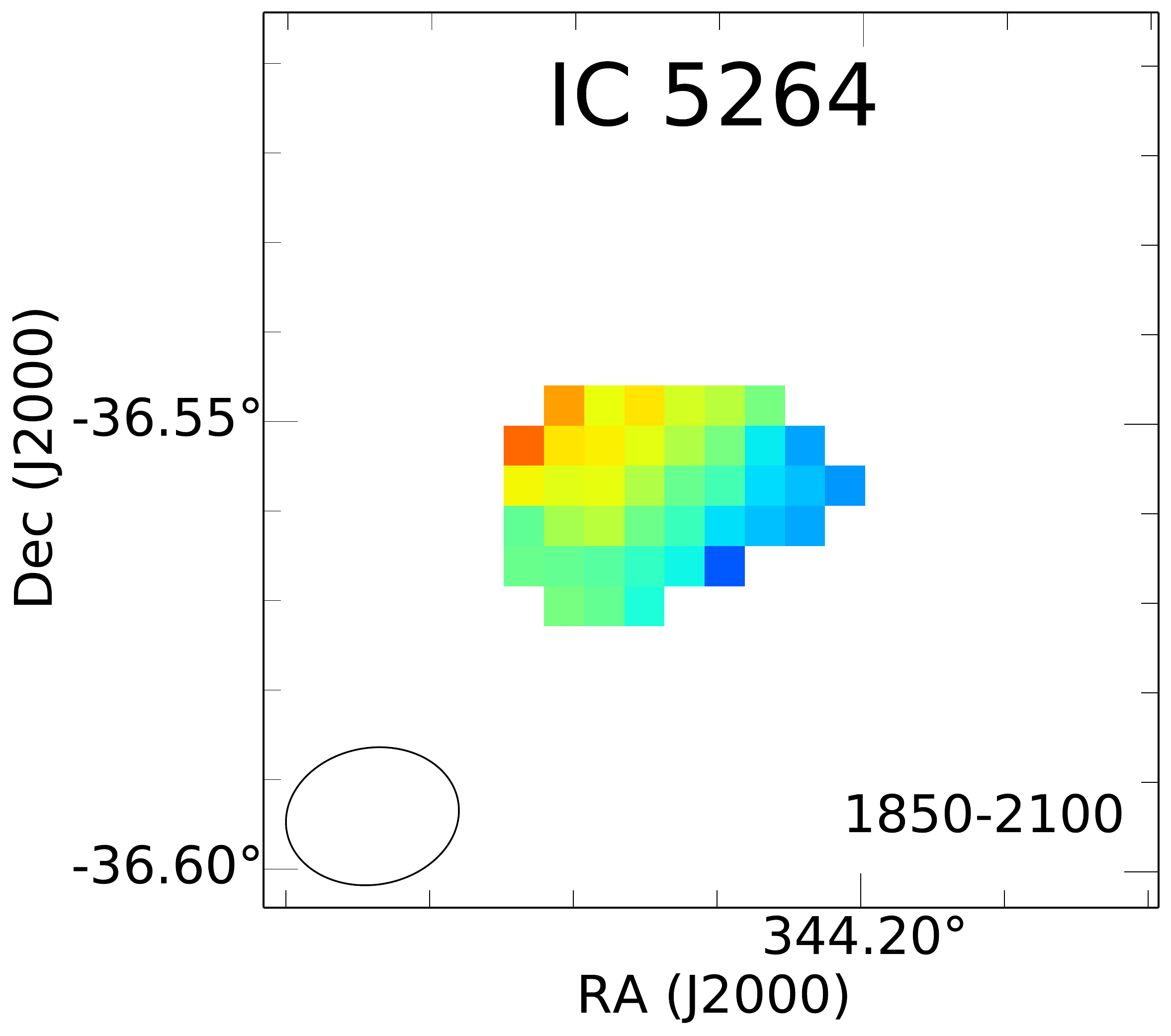}
\includegraphics[width=5cm]{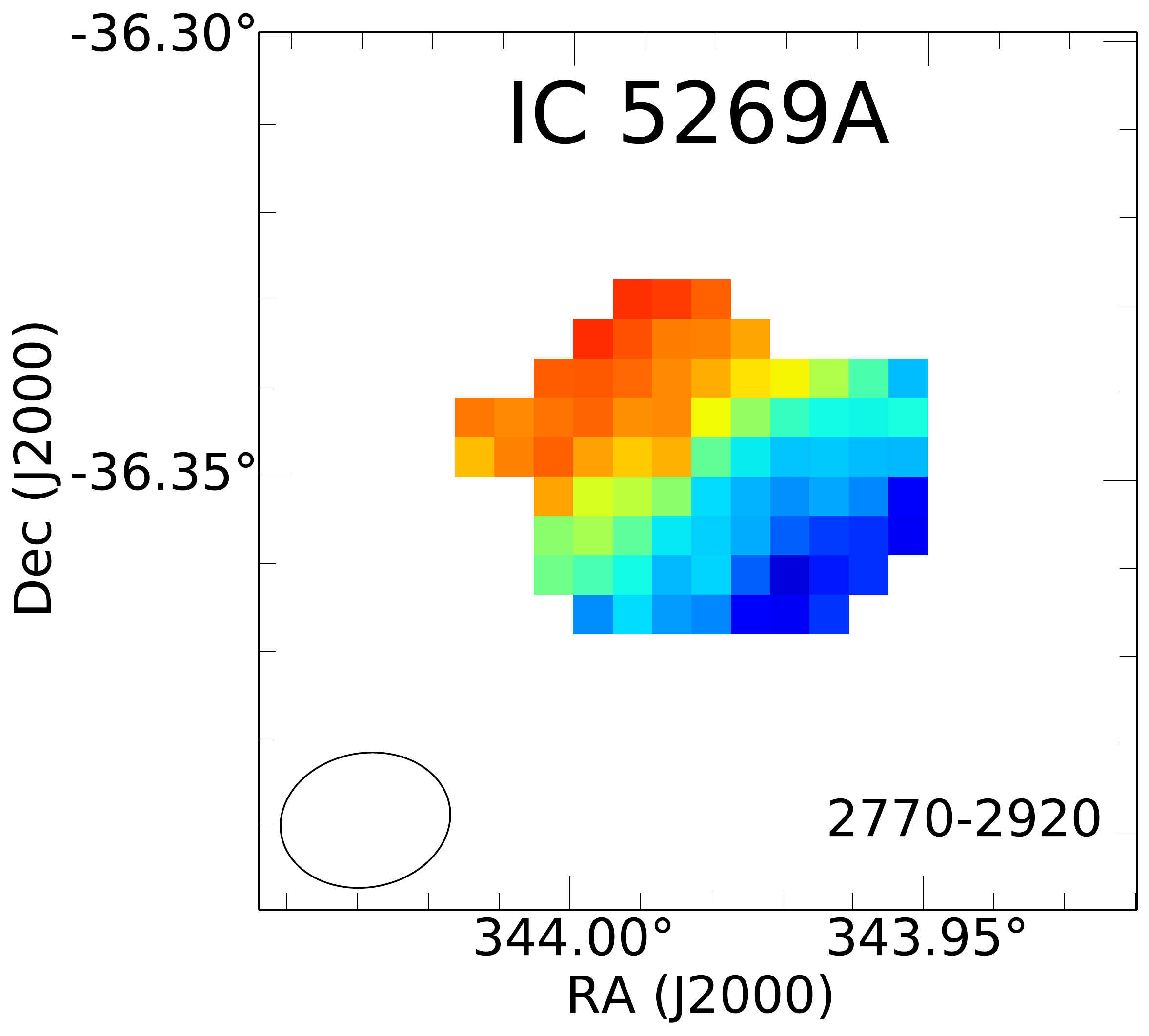}
\includegraphics[width=5cm]{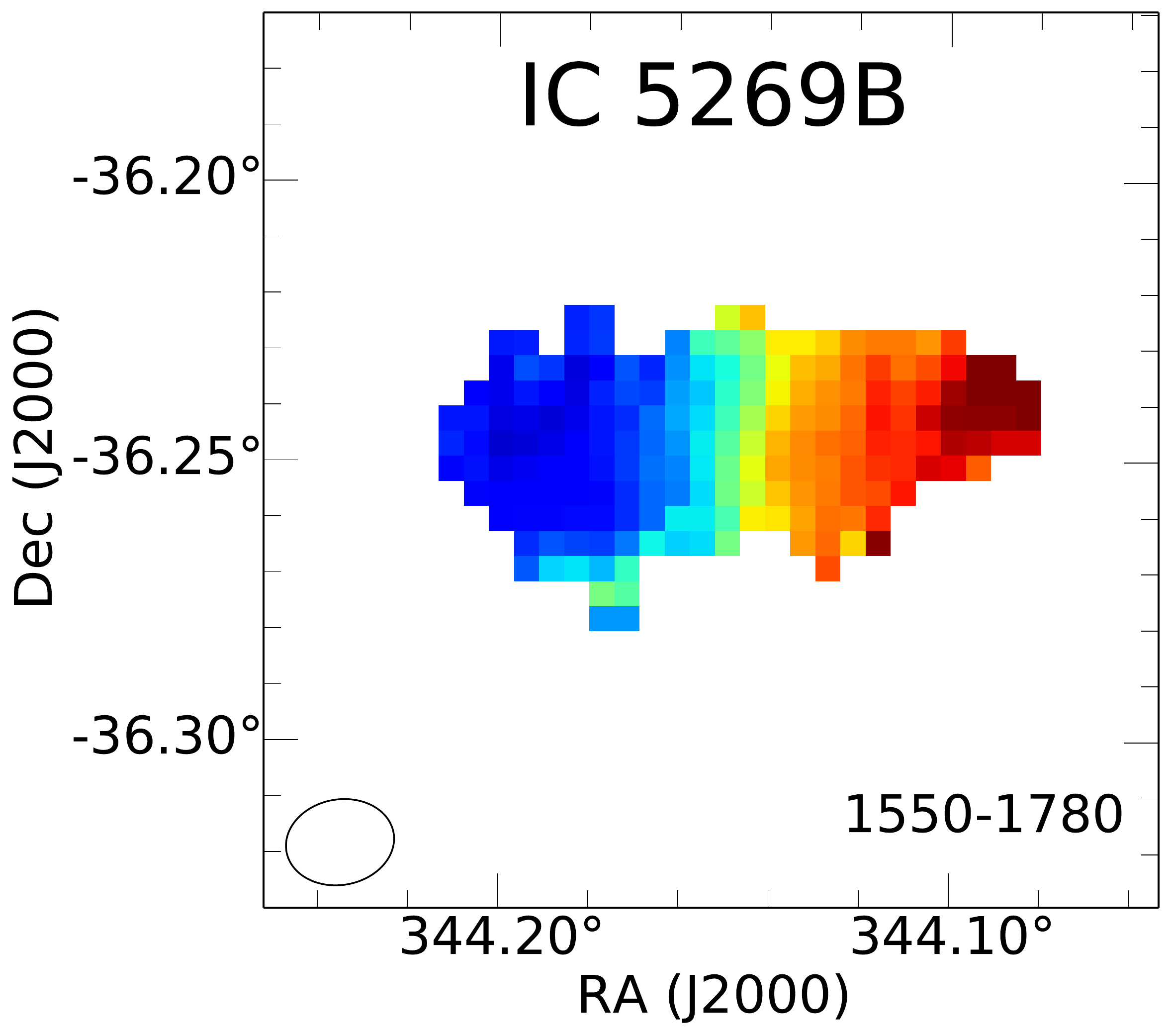}
\includegraphics[width=5cm]{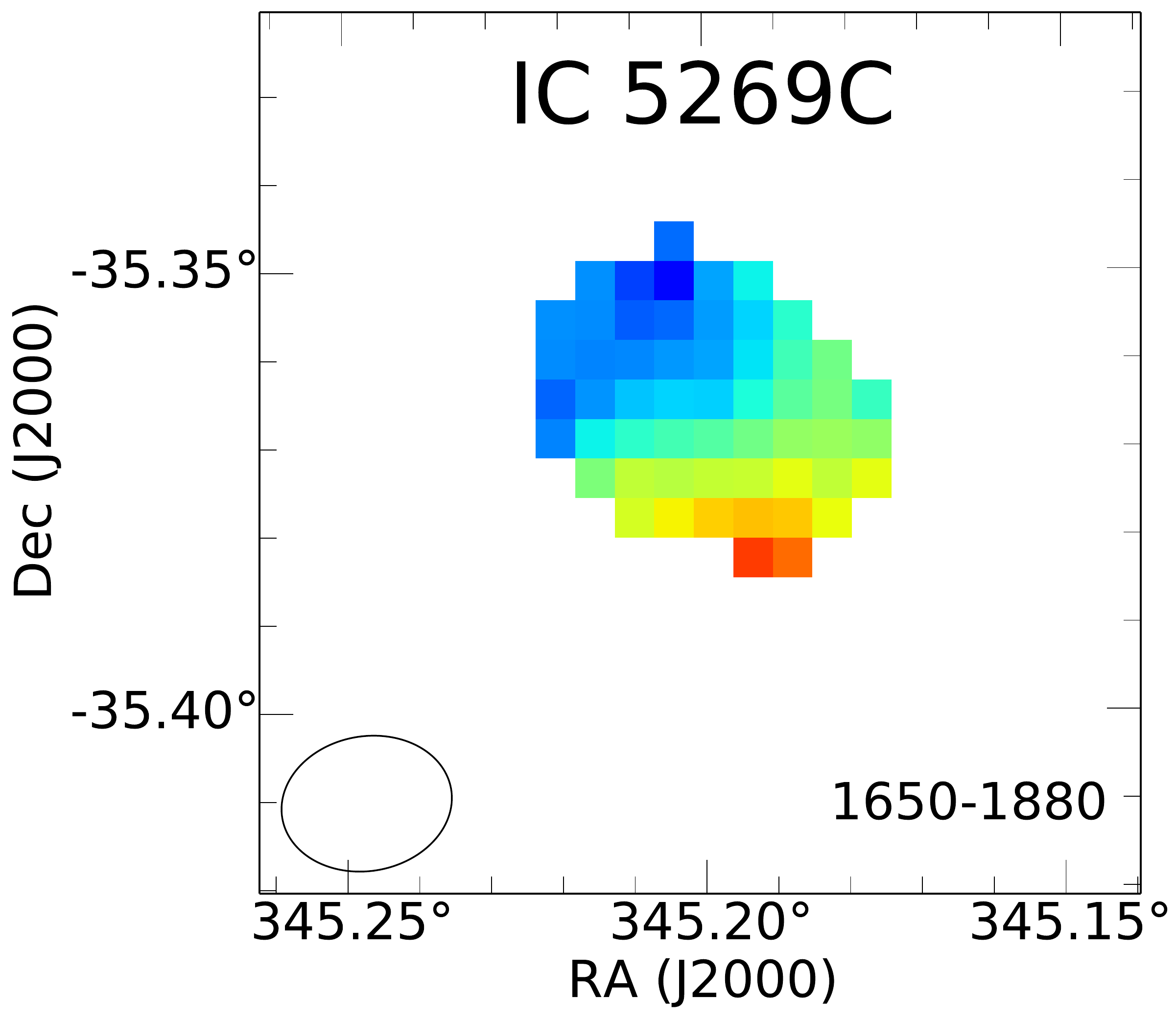}
\includegraphics[width=5cm]{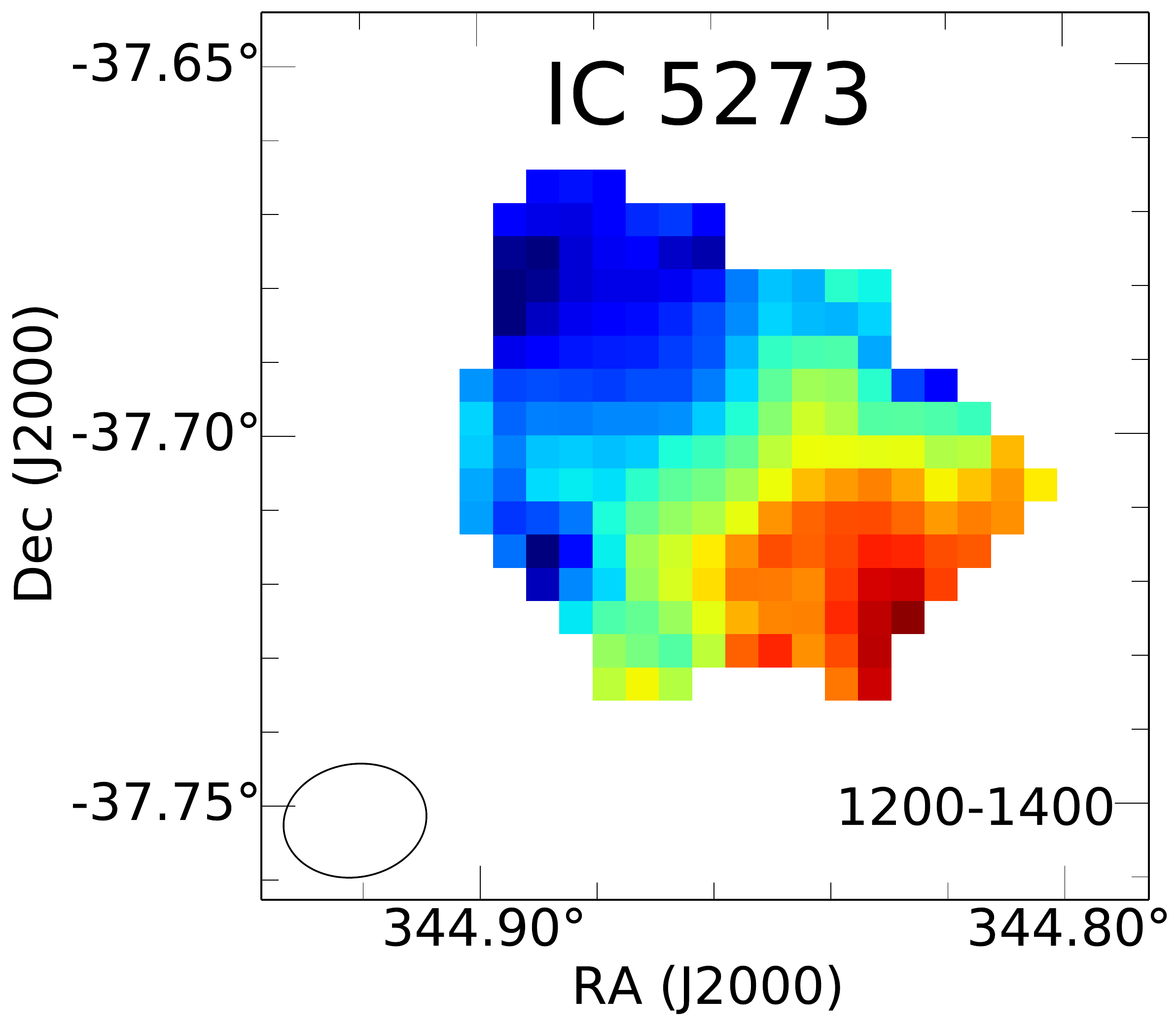}
\includegraphics[width=5cm]{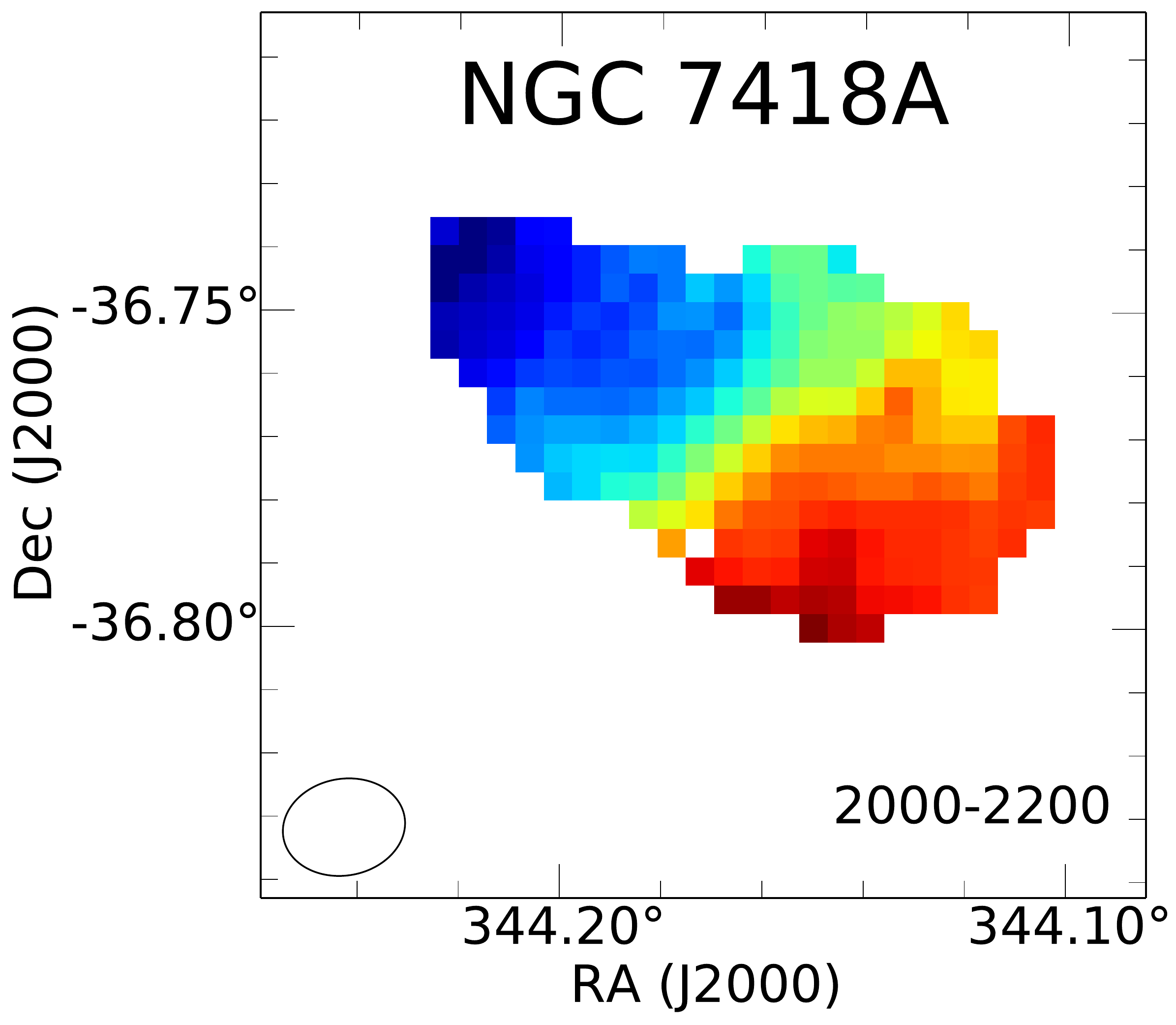}
\includegraphics[width=5cm]{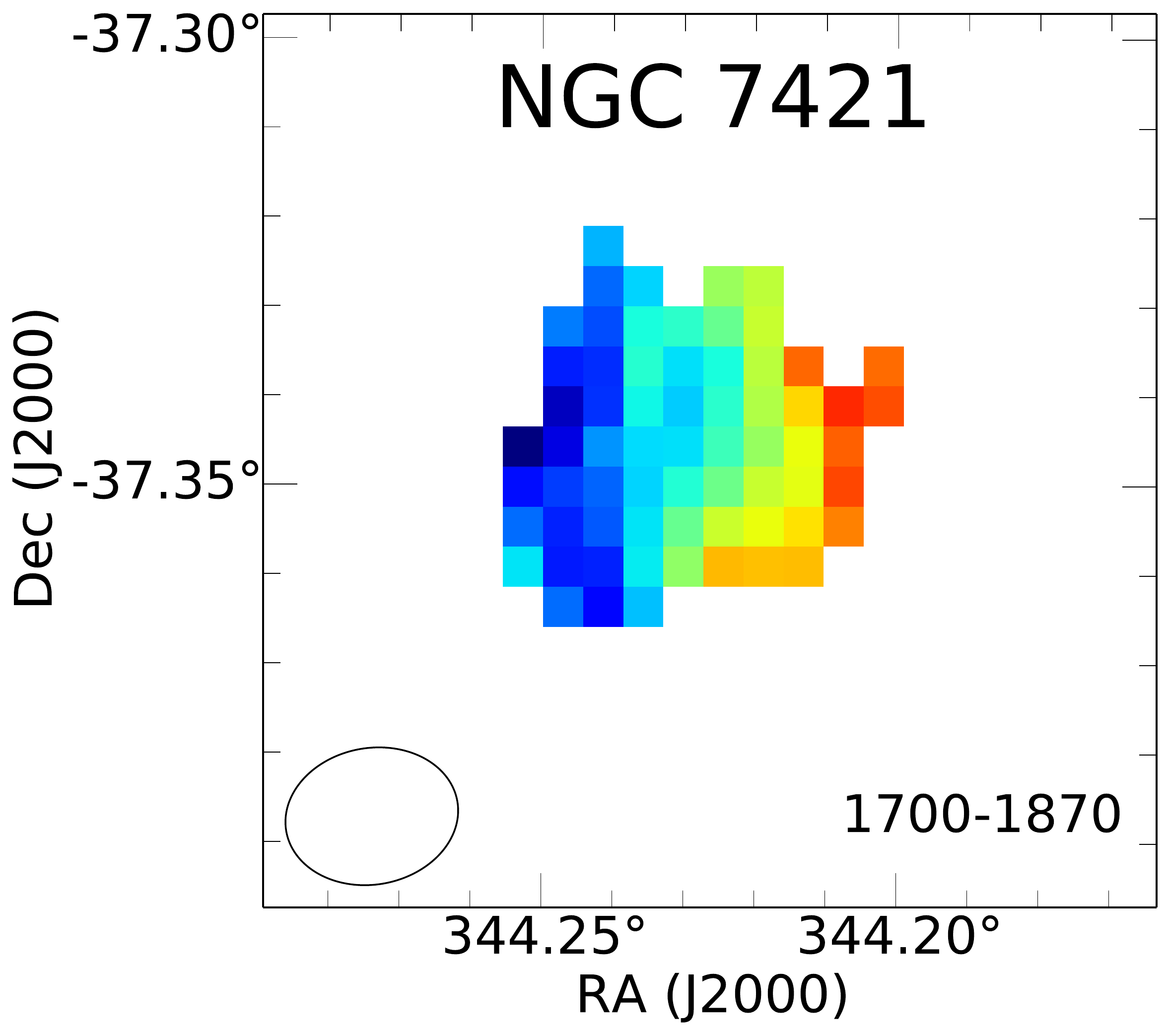}
\includegraphics[width=15cm]{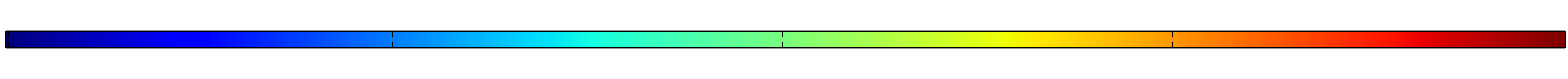}
\caption{ASKAP moment-1 \hi\ velocity field of nine of the eleven detected galaxies (for IC~5270 and NGC~7418 see Fig. \ref{fig:ic5270}). Each velocity field covers the velocity range indicated in the bottom-right in \kms\ and adopts the colour scheme represented by the horizontal colour bar. The PSF is shown in the bottom-left corner.}
\label{fig:cutouts1}
\end{figure*}

\begin{figure*}
\includegraphics[width=5cm]{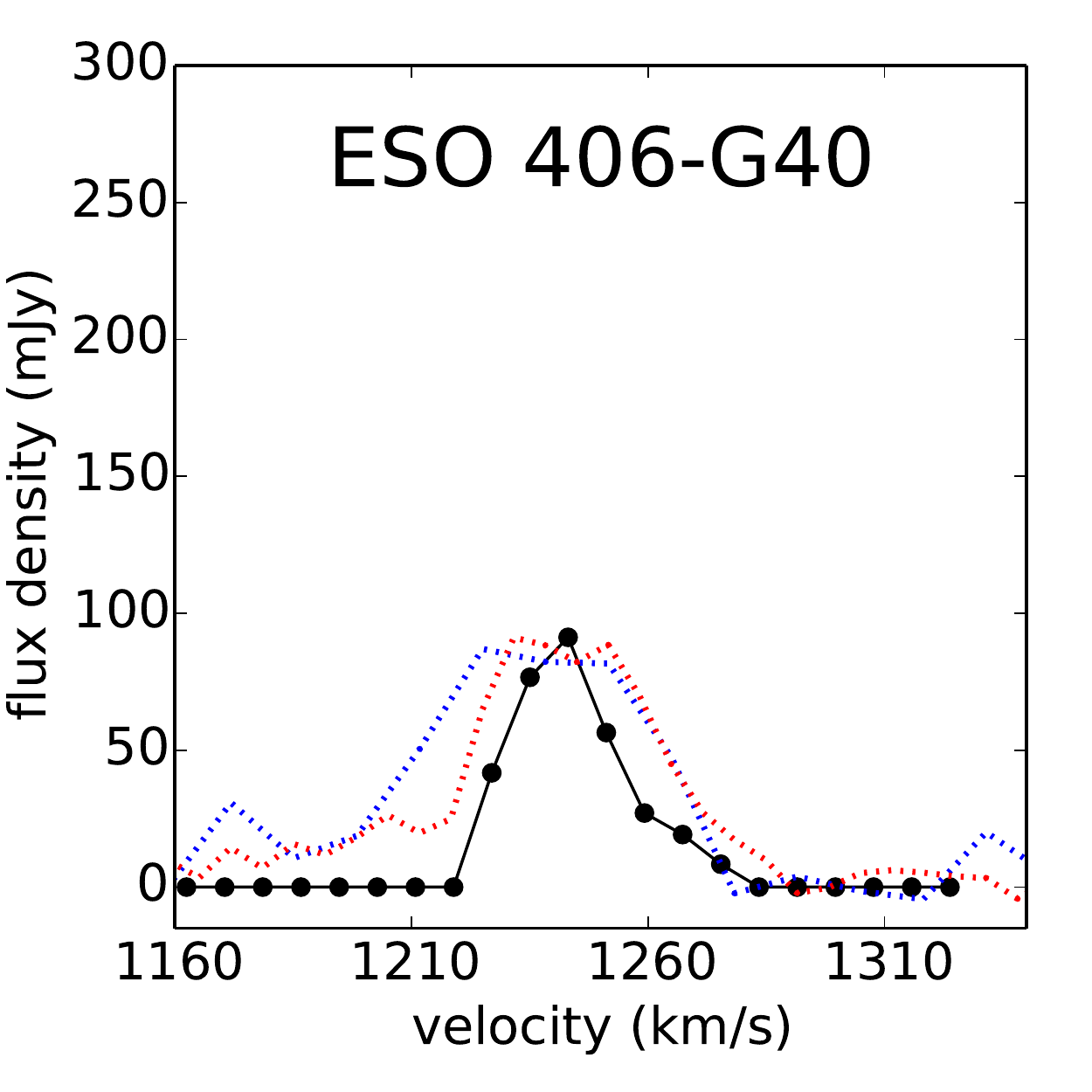}
\includegraphics[width=5cm]{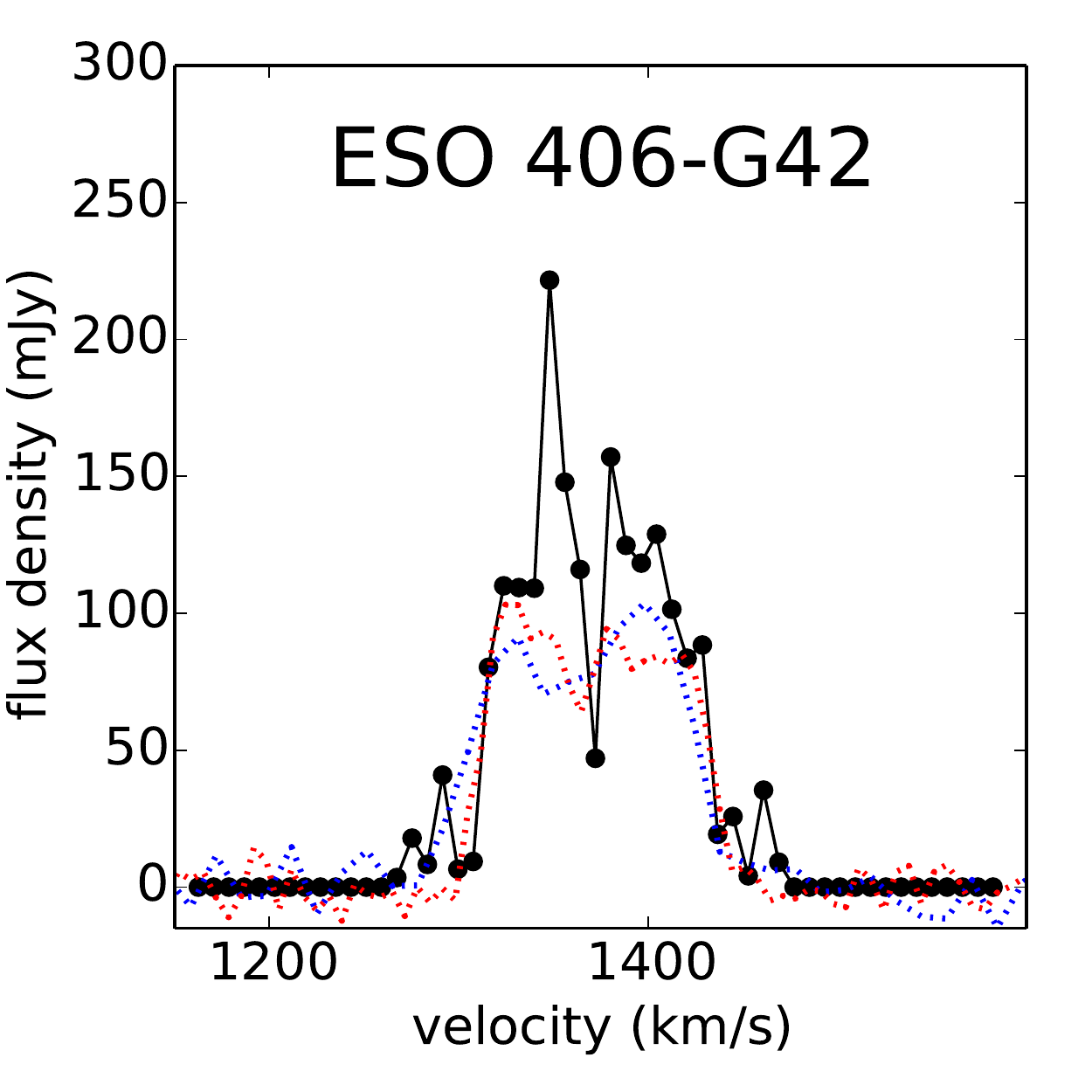}
\includegraphics[width=5cm]{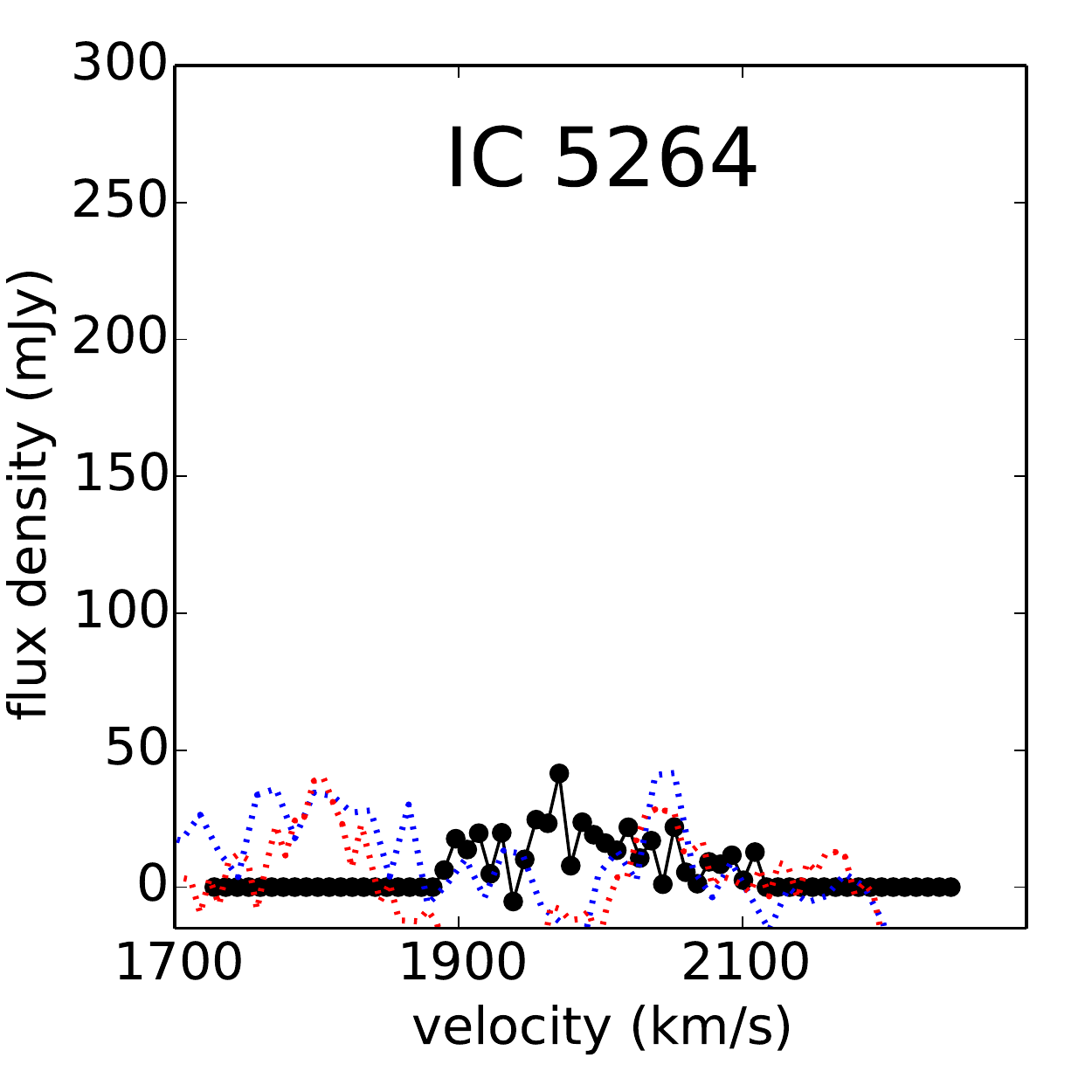}
\includegraphics[width=5cm]{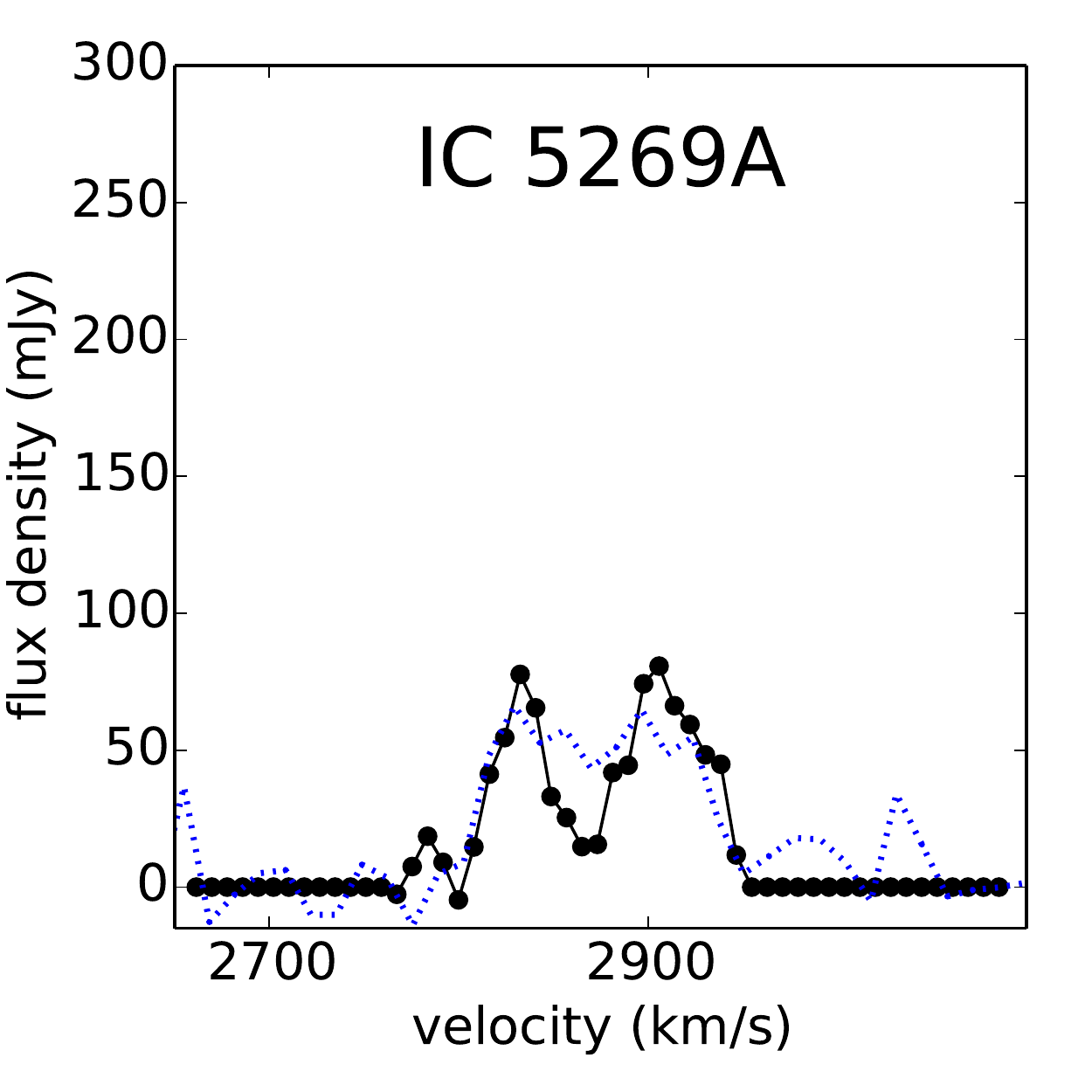}
\includegraphics[width=5cm]{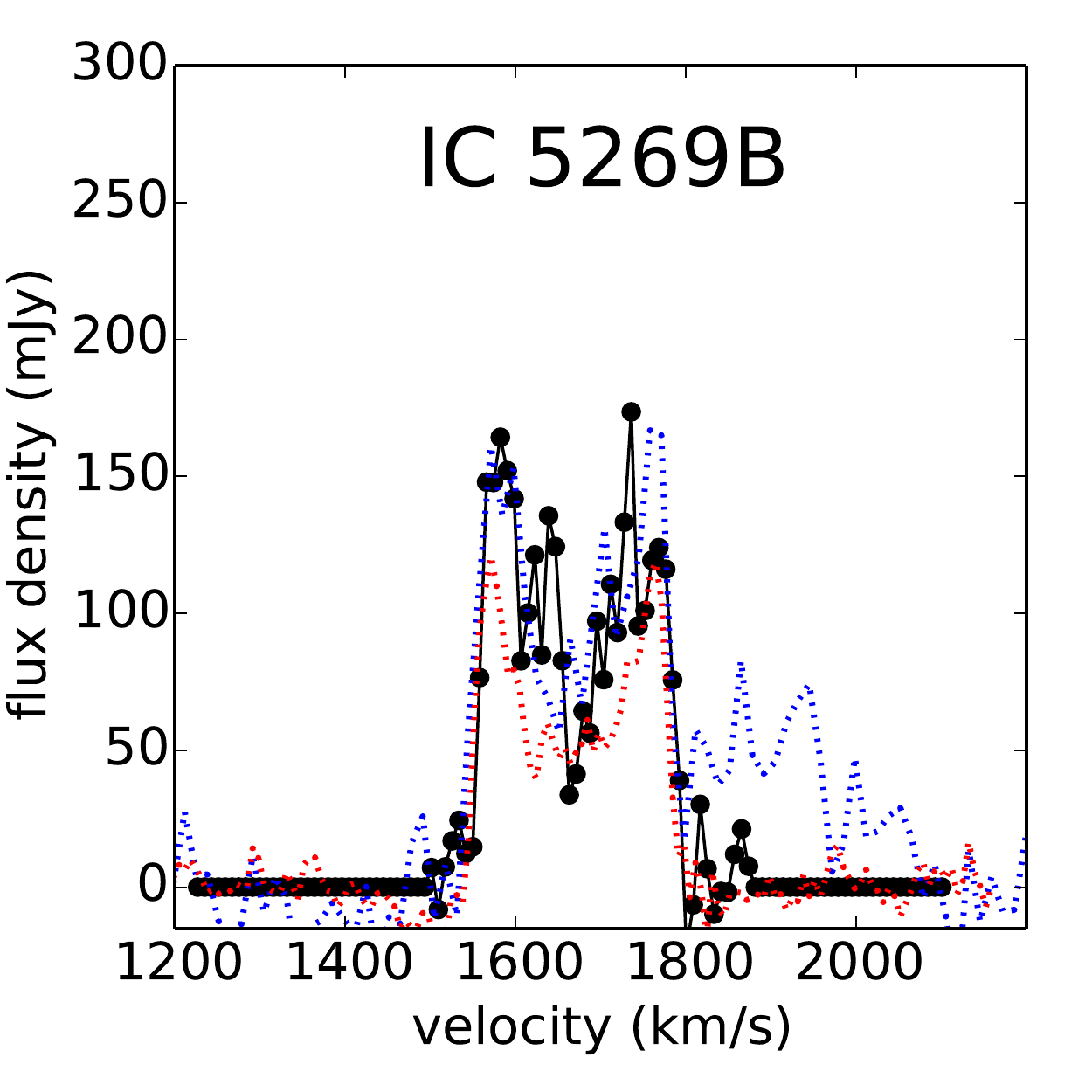}
\includegraphics[width=5cm]{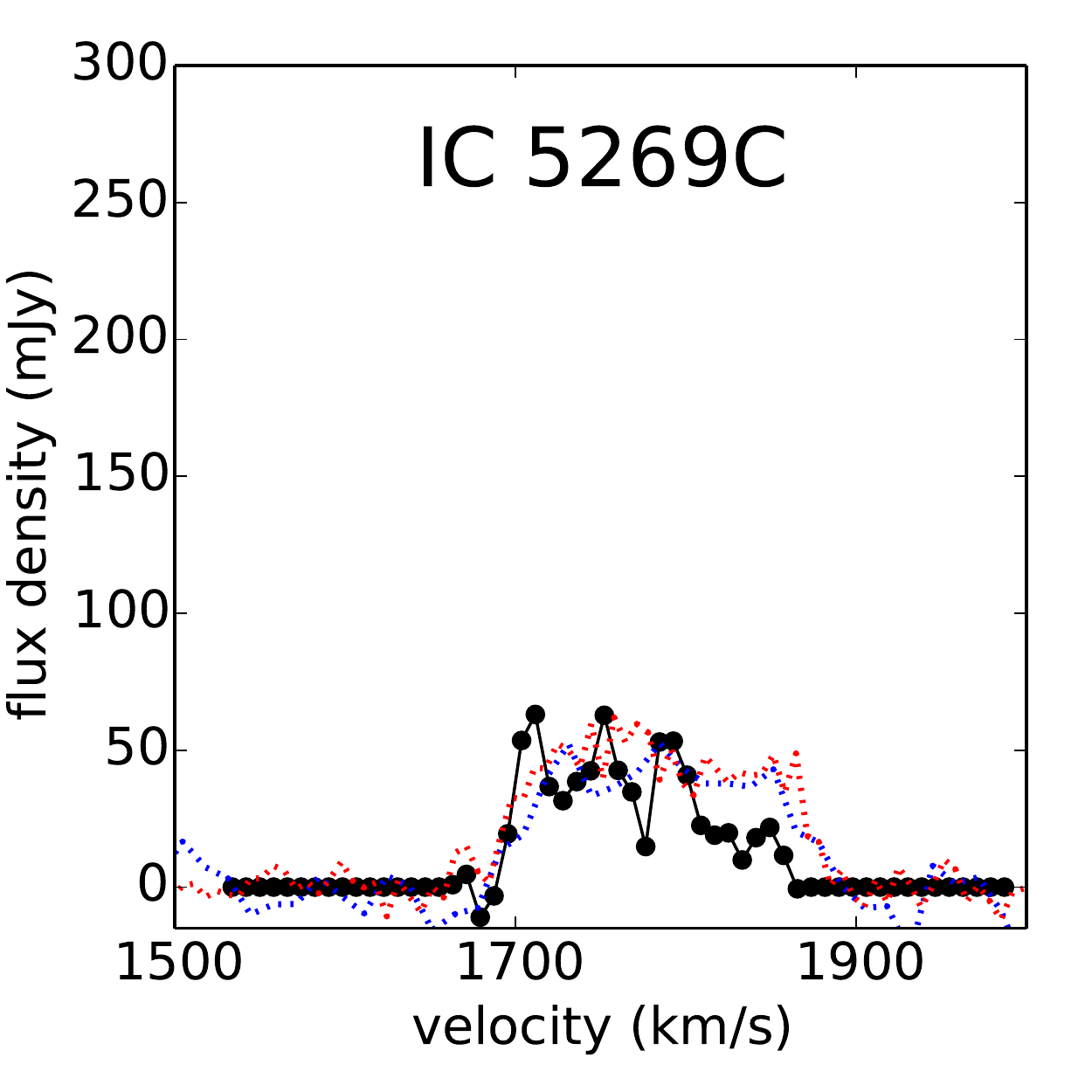}
\includegraphics[width=5cm]{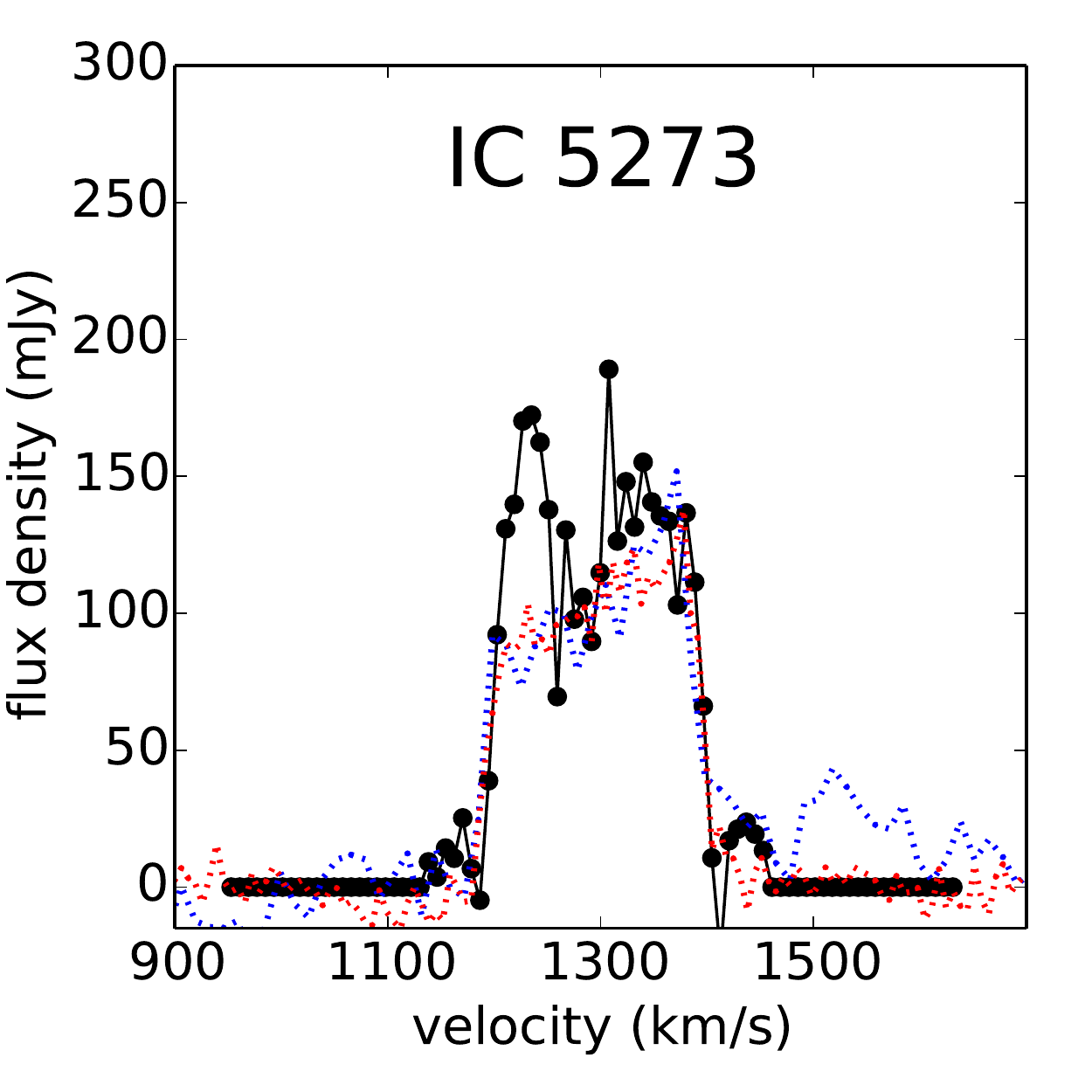}
\includegraphics[width=5cm]{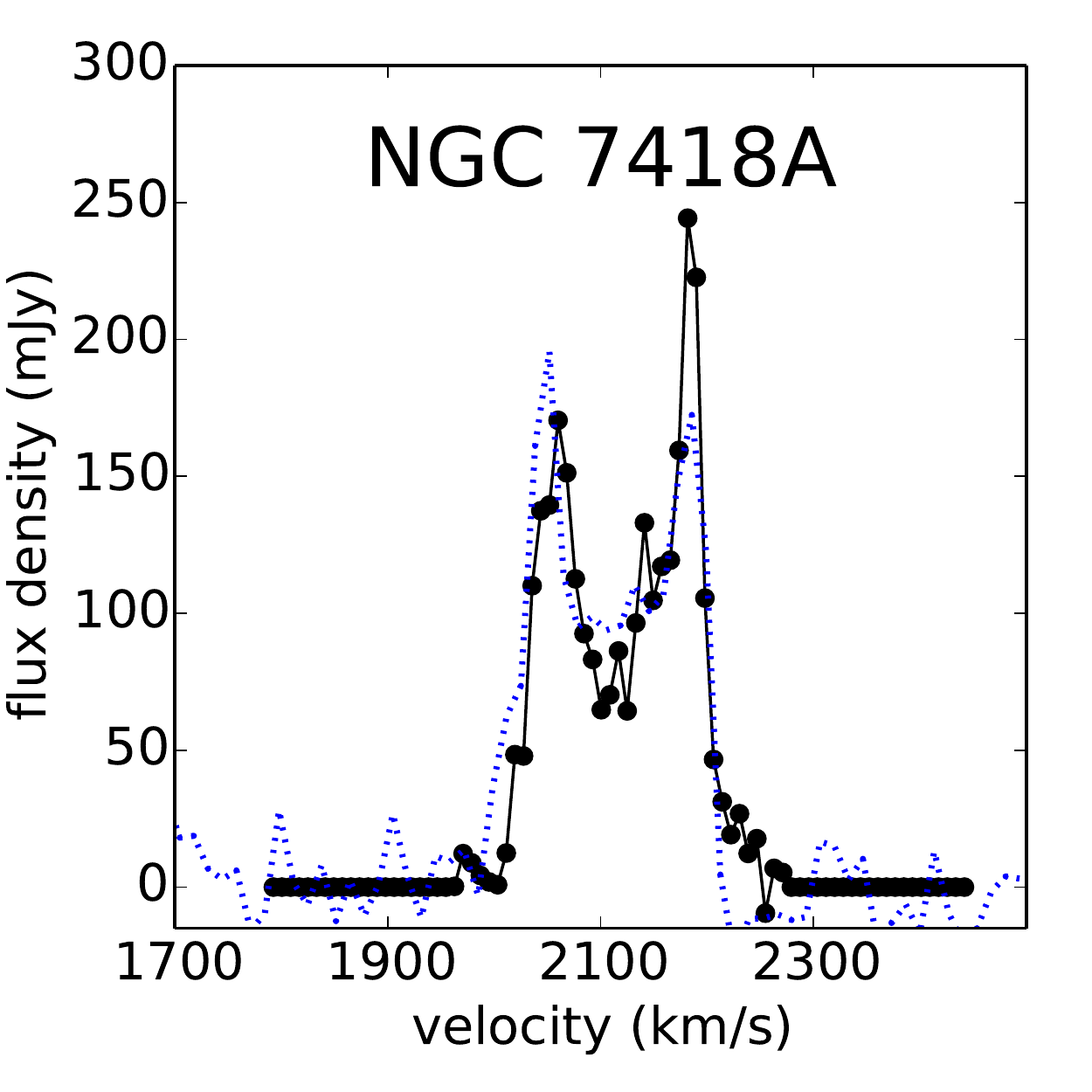}
\includegraphics[width=5cm]{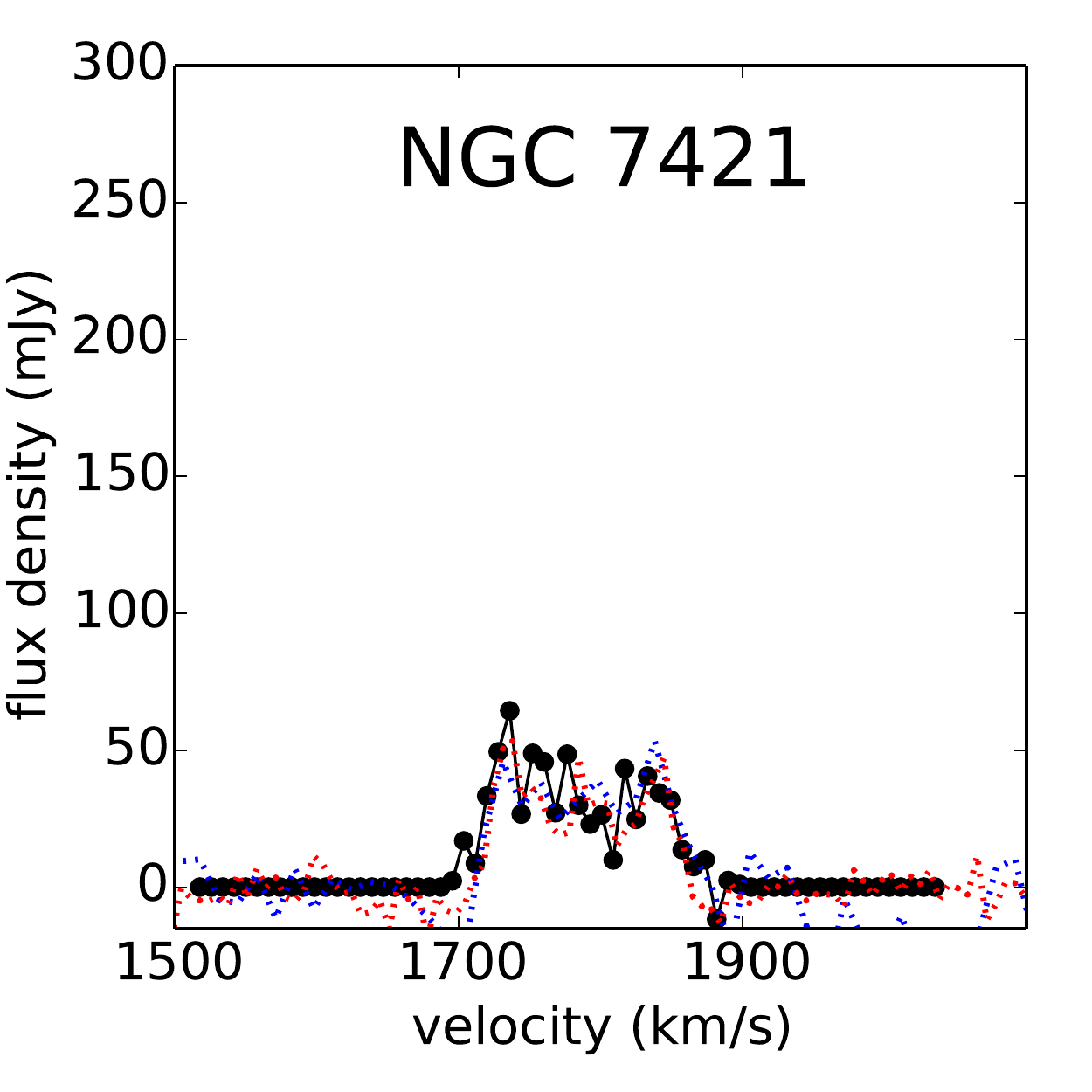}
\caption{ASKAP integrated \hi\ spectrum (black dots and solid line) of nine of the eleven detected galaxies (for IC~5270 and NGC~7418 see Fig. \ref{fig:ic5270}). The spectrum is constructed using only voxels included in the mask generated by \sofia\ and used to make the \hi\ image and velocity field in Figs. \ref{fig:cutouts0} and \ref{fig:cutouts1}, respectively. For comparison, HIPASS spectra are shown with a blue dotted line, and spectra by Kilborn et al. (2009) are shown with a red dotted line.}
\label{fig:cutouts2}
\end{figure*}

\section{\hi\ in the IC~1459 group}
\label{sec:results}

\subsection{\hi\ source finding}
\label{sec:sofia}

We search for \hi\ in the final mosaic cube using the \sofia\ source-finding package \citep{serra2015}. In order to take into account noise variations within the \hi\ cube, we provide an input weights image equal to the inverse of the noise image obtained with Eq. \ref{eq2}, and we let \sofia\ remove noise variations along the frequency axis. We use the smooth+clip source detection algorithm, which smooths the input (weighted) cube with a set of $N$ filters and detects emission above the required threshold after each smoothing step. In this case we smooth the cube on the sky and/or in velocity using, respectively, a Gaussian filter of FWHM = 3 pixels (the pixel size is 15 arcsec) and a box filter of width 3, 5, 9, 15 and 23 channels (the channel width is 8 \kms), and adopt a detection threshold of $4\sigma$.

We merge detected voxels into objects  using a merging radius of 1 pixel and 3 channels. We reject objects with linear size smaller than 3 pixels and spanning less than 5 channels, and apply the \cite{serra2012b} algorithm to reject detections less reliable than 99 percent. Finally, we dilate the mask of all reliable detections to recover their total \hi\ flux. Mask dilation stops if the total source flux grows by less than 2 percent, resulting in a typical dilation of 2 pixels in each channel.

All the above processing is performed within \sofia. We refer the reader to the paper describing this software package \citep{serra2015} for details on its working.

\subsection{Mass, morphology and kinematics of the detected \hi}

We detect \hi\ in eleven galaxies. We show all detections in Fig. \ref{fig:mosaic}, where red contours represent \hi\ at a column density of $10^{20}$ cm$^{-2}$. As shown in the figure, we detect two \hi\ clouds near the disc of IC~5270. The clouds are within a few arcmin north, and in the same recessional velocity range, of IC~5270. We detect an \hi\ cloud at the north-west edge of NGC~7418, too. These two galaxies are of particular interest and we discuss them in detail in Sec. \ref{sec:ic5270}. Here we describe the other nine objects.

We show the \hi\ images, velocity fields and integrated spectra of these nine detections in Figs. \ref{fig:cutouts0}, \ref{fig:cutouts1} and \ref{fig:cutouts2} respectively. In the latter figure we also compare the ASKAP spectra to those obtained using the Parkes telescope as part of HIPASS \citep{barnes2001} and by \cite{kilborn2009}. Overall, the agreement between the ASKAP and Parkes spectra is good. This is confirmed in Table \ref{tab:memb}, where we compare the \hi\ masses measured from the different datasets. The uncertainty on the ASKAP \hi\ mass is obtained by summing in quadrature the 20 percent flux uncertainty caused by our limited knowledge of the formed beams' shape (Sec. \ref{sec:bpgain}) and the uncertainty related to the presence of noise in the \hi\ cube. For each detection, we estimate the latter term by placing the 3D mask produced by \sofia\ at $\sim100$ random positions in the mosaic cube, and measuring the scatter in the resulting distribution of \hi\ mass values. This is typically $\sim10$ percent of the source \hi\ mass. The agreement between the ASKAP and Parkes \hi\ masses is good within the errors.

As shown in Table \ref{tab:memb}, we fail to detect one previously known \hi\ source: DUKST~406-83. The total \hi\ flux and line width of this galaxy are $\sim1.6$ Jy \kms\ and $\sim80$ \kms, respectively \citep{kilborn2009}. This would imply an average flux density of $\sim20$ \mJybeam\ across the \hi\ profile ($\sim2\sigma$ per channel over ten channels in our cube) if the source is unresolved by the PSF of our \hi\ cube. If the emission is resolved on the sky (within individual channels) the signal-to-noise ratio will be even lower. Visual inspection of the ASKAP \hi\ cube at the location of DUKST~406-83 shows emission at the $\sim2\sigma$ level in 4 consecutive channels at the galaxy's recessional velocity -- too faint to be detected reliably with our source-finding strategy (Sec. \ref{sec:sofia}).

Excluding IC~5270 and NGC~7418 -- which we discuss in Sec. \ref{sec:ic5270} -- the brightest sources in the ASKAP \hi\ mosaic cube are ESO~406-G42, IC~5269B, IC~5273 and NGC~7418A. These objects have an \hi\ mass of a few times $10^9$ \msun\ and a fairly typical \hi\ spectrum. Among them, ESO~406-G42 stands out as one in relatively poor agreement with the Parkes spectra. Its \hi\ velocity field appears noisy, too. This is explained by the position of this galaxy close to the edge of the observed field, in a region where the noise is $\sim50$ percent higher than in the inner region of the mosaic. Some discrepancy between ASKAP and Parkes spectra can also be seen at the high-velocity end of the \hi\ spectra of IC~5269B and IC~5273. In IC~5269B the emission at $\sim1900$ \kms\ recorded by the HIPASS spectrum is caused by confusion with \hi\ in IC~5270. This emission is not visible in the spectrum from \cite{kilborn2009} because they use a smaller aperture. In IC~5273 visual inspection of the HIPASS cube suggests that the emission at $\sim1500$ \kms\ is likely to be a noise peak, possibly combined with a non-flat spectral baseline. This emission is not detected in the spectrum of \cite{kilborn2009}.

Other galaxies in Figs. \ref{fig:cutouts0} to \ref{fig:cutouts2} have an \hi\ mass of $10^9$ \msun\ or less, and significantly lower S/N. This is reflected in the quality of their \hi\ images and velocity fields. Yet, the agreement with the Parkes spectra is good even for very faint objects such as ESO~406-G40, IC~5269A, IC~5269C and NGC~7421. Some slight differences are visible for ESO~406-G40 and  IC~5269C. In the former the ASKAP spectrum is narrower than the Parkes spectra. Following visual inspection of the ASKAP cube, the most probable explanation is that the ASKAP data miss some low column density \hi\ emission at velocities away from systemic. An additional effect is that the ASKAP velocity resolution is $\sim2$ times better than that of the Parkes data. This would be noticeable in particular for a narrow spectrum such as that of ESO~406-G40. In IC~5269C the optical disc is warped, and we seem to be missing the low-column-density, red-shifted, southern \hi\ end of the warp. This galaxy is close to the edge of the mosaic, where the noise is higher than in the inner regions.

The most extreme case among the faint \hi\ sources is IC~5264, as can be seen from the integrated spectrum in Fig. \ref{fig:cutouts2}. This galaxy is nearly edge-on and, therefore, the low \hi\ flux is spread over a large velocity range, $\sim250$ \kms\ in our data. As a result, the peak \hi\ emission in the ASKAP cube is $<4\sigma$. The ASKAP \hi\ velocity range of IC~5264 is consistent with that shown by \citet[][$\sim300$ \kms]{walsh1990} and with the \hi\ spectrum that we obtain by reducing and imaging the ATCA data of \citet[][ATCA project code C530]{oosterloo1999}. The detection of such faint, edge-on objects in future large surveys will require the use of multi-scale automated source-finding similar to the one used here (Sec. \ref{sec:sofia}), as well as \hi\ cubes relatively free from artefacts.

Our data show hints of a disturbed \hi\ morphology in NGC~7418A, whose asymmetric \hi\ disc follows the shape of the low surface brightness, extended, clumpy optical star-forming disc with prominent spiral arms. \cite{thilker2007} classify it is as ``mixed-type extended UV disc'', i.e., an object exhibiting ``a large, blue low-surface-brightness zone and structured UV-bright complexes at extreme galactocentric distances beyond the traditional star-formation threshold''. The proximity of NGC~7418A to the (also disturbed) early-type central galaxy IC~1459 suggests that the two may have interacted recently. The intra-group \hi\ detected by \cite{walsh1990} and \cite{oosterloo1999} north of NGC~7418A may have been removed from this galaxy during such interaction. Alternatively, NGC~7418A may have recently accreted gas from this diffuse structure, and the faint, clumpy star formation in its outer disc may have been triggered by such accretion.

Another galaxy that may be undergoing some interaction with the group environment is IC~5273. Its \hi\ morphology is suggestive of gas near the disc edge being pushed north-west relative to the stellar body. However, the signal in the gas disc outer regions is very low, and deeper, higher-resolution data would be needed to clarify the situation. Whether or not IC~5273 is losing some of its gas to the intra-group medium, we note that some ram-pressure stripping is definitely happening in this group, as demonstrated by \cite{ryder1997} for NGC~7421. Their ATCA data show that \hi\ at a column density below $\sim10^{20}$ cm$^{-2}$ is clearly being pushed off the galaxy in a eastward direction, consistent with the compressed, star-forming west side of the optical disc (Fig. \ref{fig:cutouts0}). The column density sensitivity of our data is not sufficient to study this object in more detail, but the slight offset between \hi\ and optical emission visible in Fig. \ref{fig:cutouts0} is in agreement with the deeper ATCA image.

Finally, the other bright and well-resolved galaxy in our data, IC~5269B, does not show any clear sign of a disturbed \hi\ morphology; and the slight offset between \hi\ and stellar body in ESO~406-G42 and IC~5264 is most likely due to the low S/N.

\subsection{\hi\ clouds in IC~5270 and NGC~7418}
\label{sec:ic5270}

\begin{figure*}
\includegraphics[height=5cm]{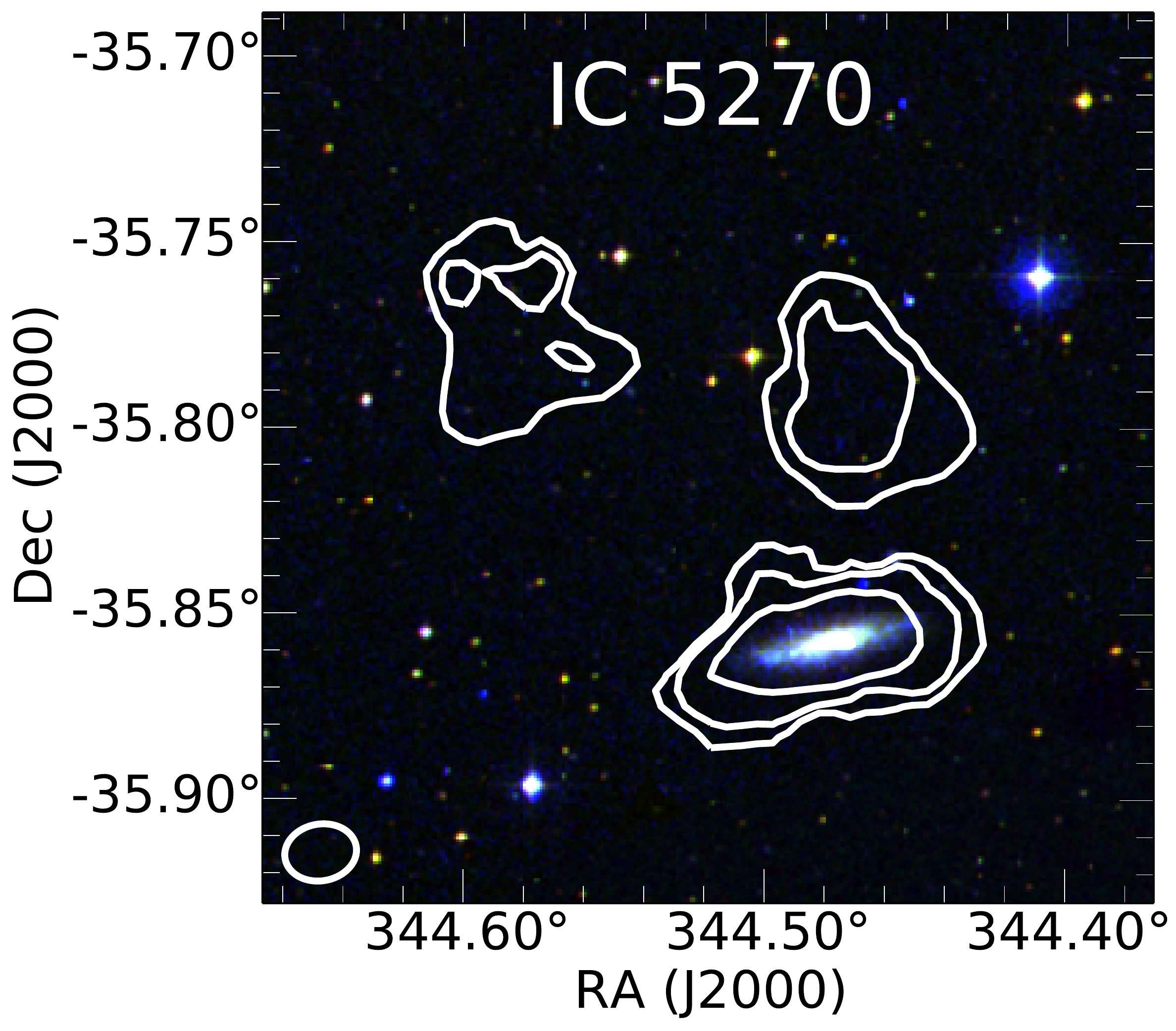}
\includegraphics[height=5cm]{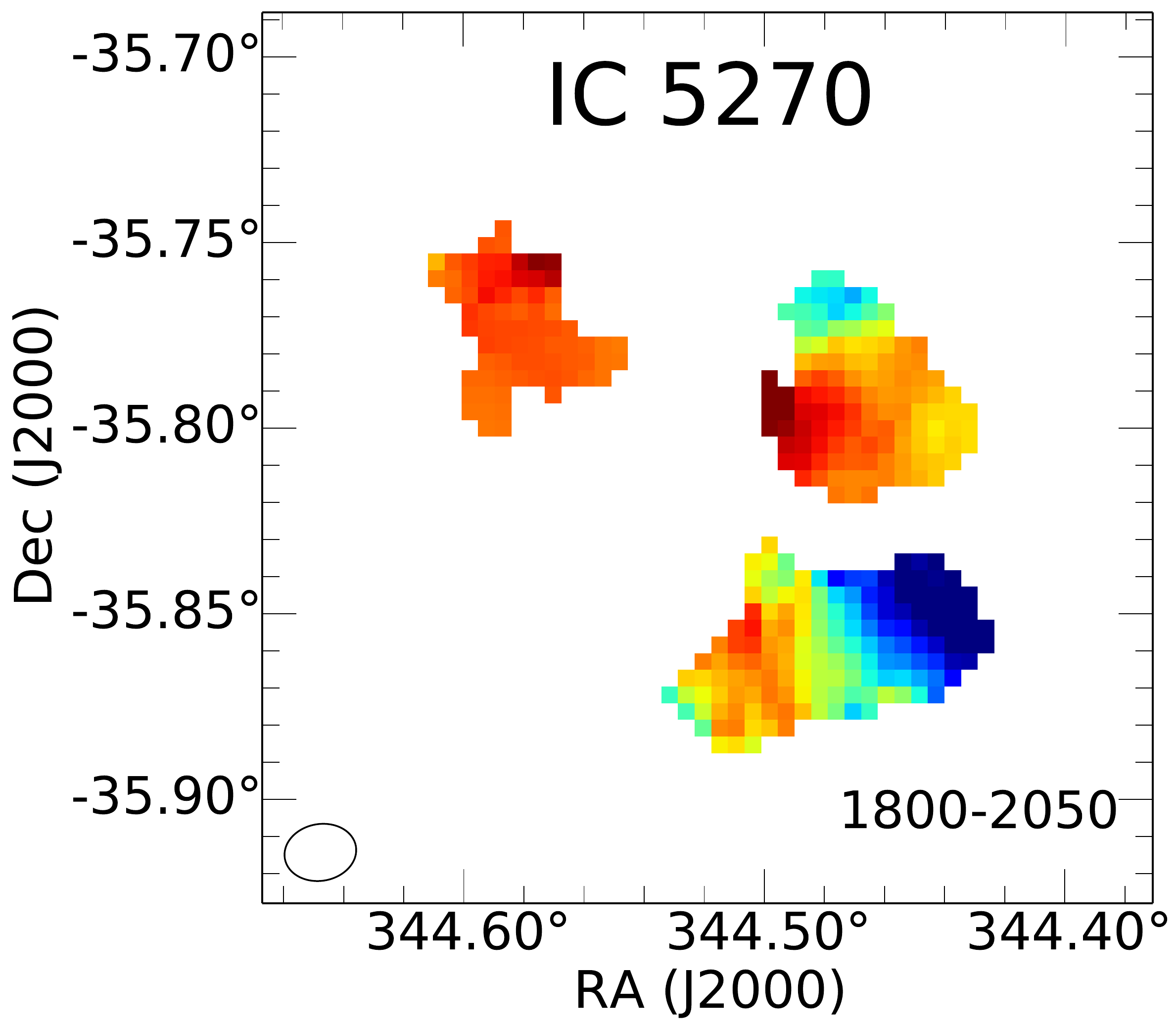}
\includegraphics[height=5cm]{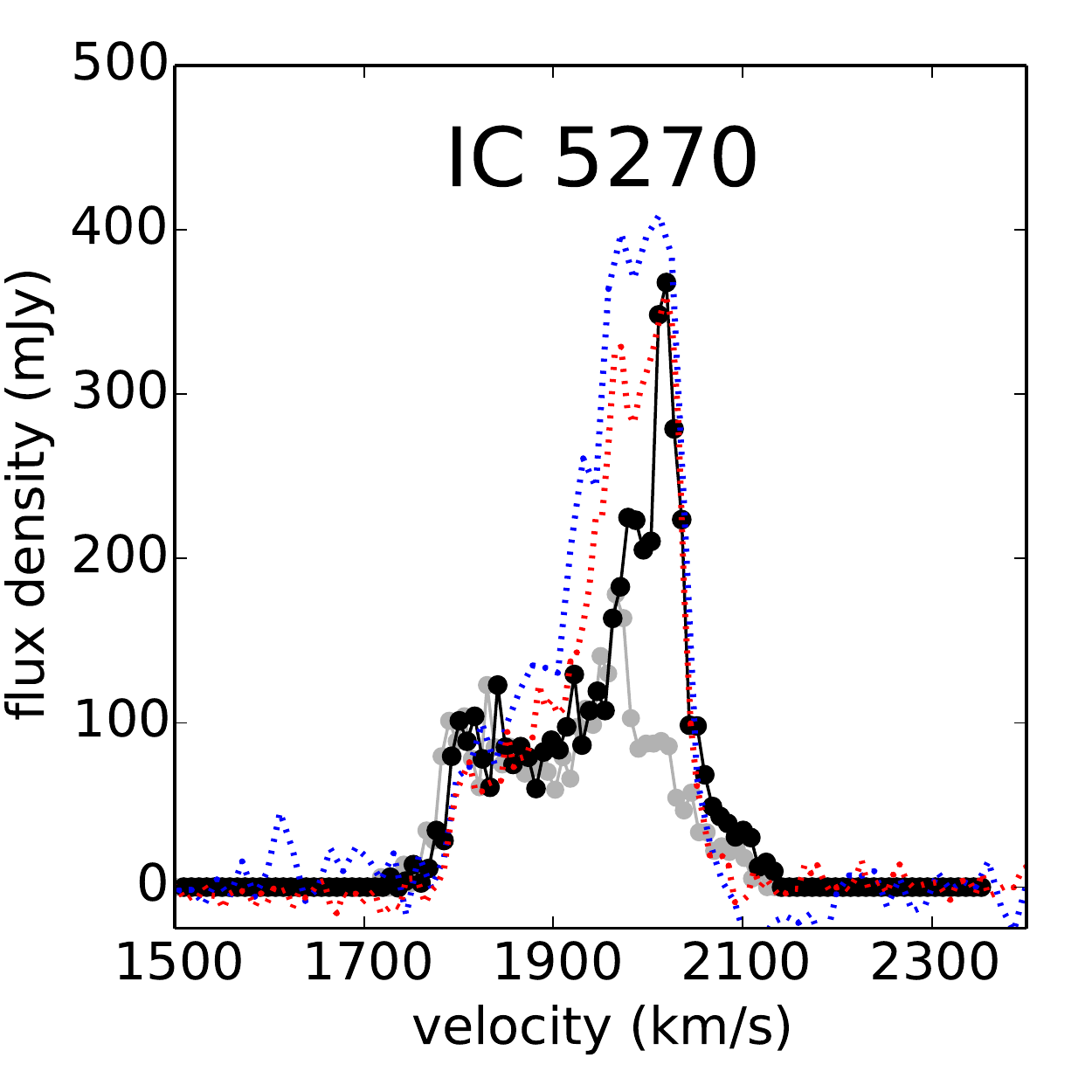}
\includegraphics[height=5cm]{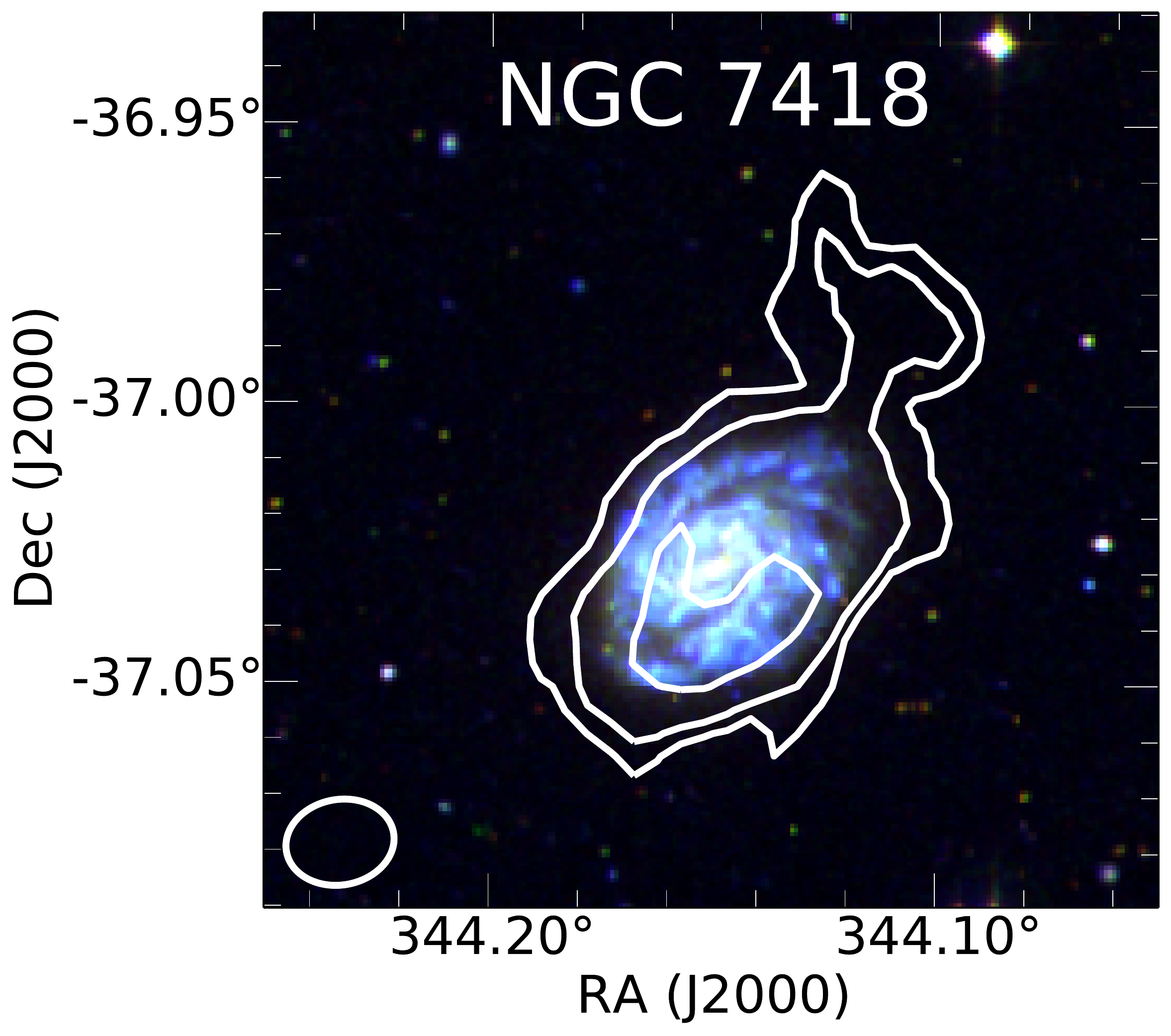}
\includegraphics[height=5cm]{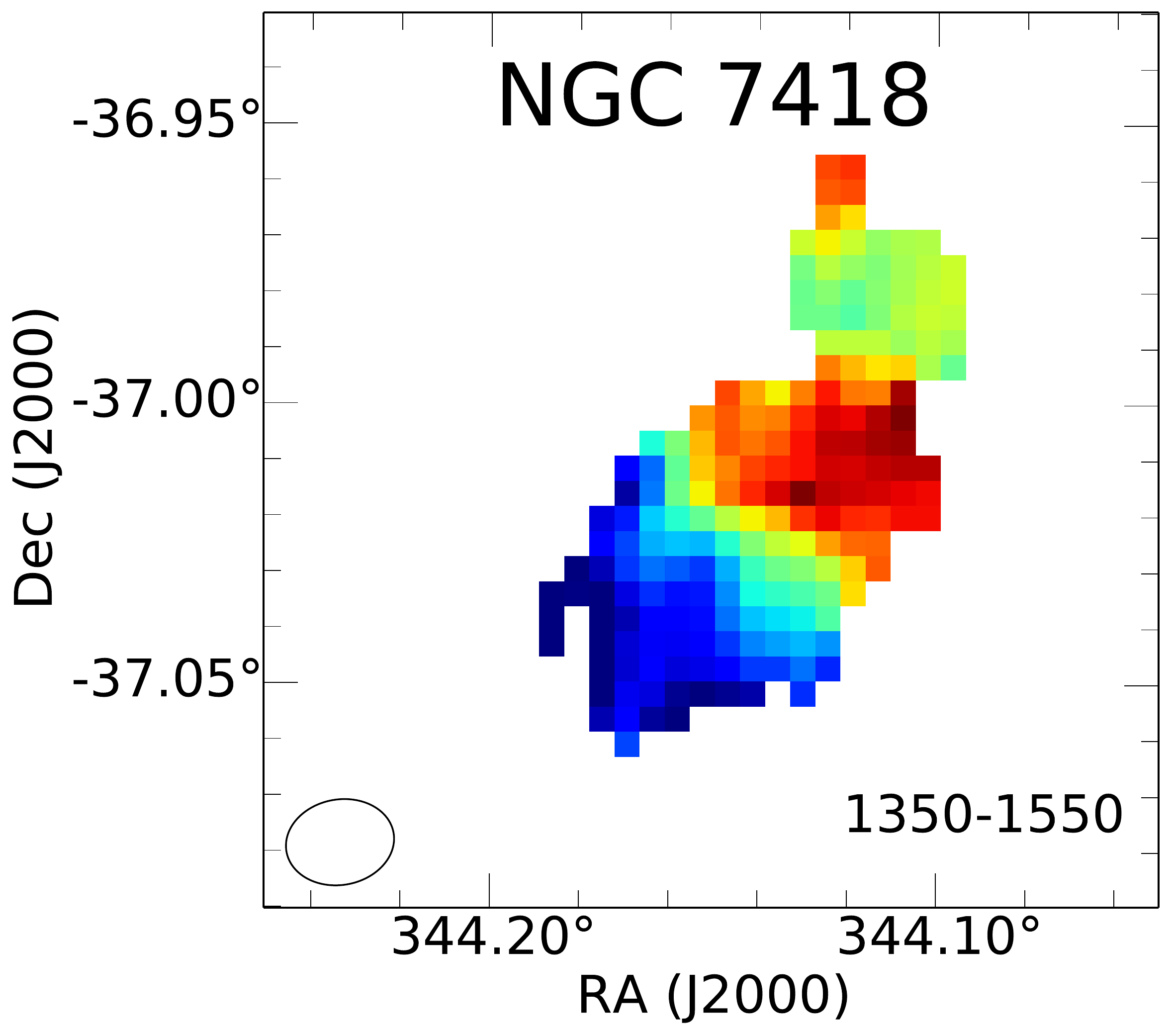}
\includegraphics[height=5cm]{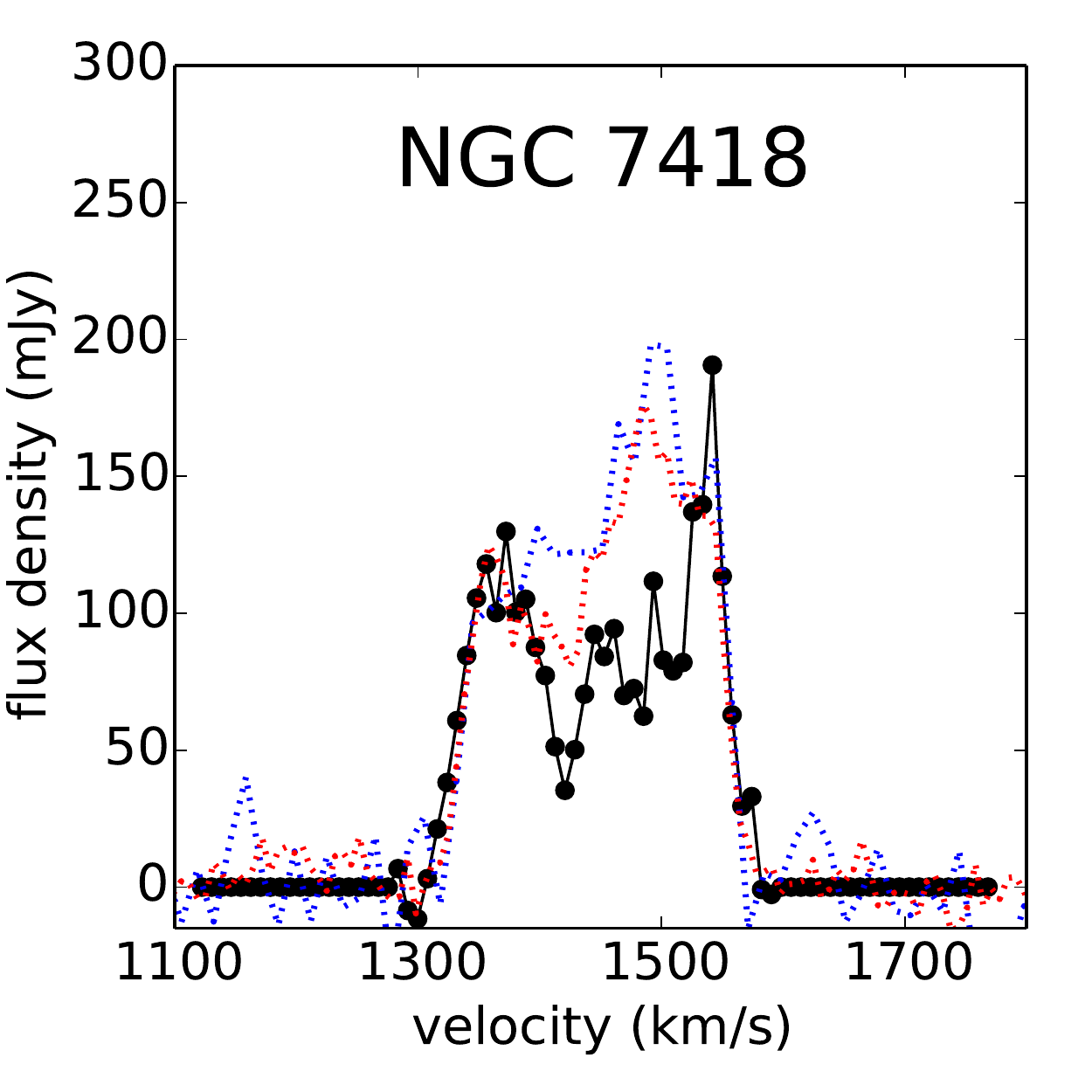}
\caption{ASKAP \hi\ contours (white) overlaid on an RGB image, \hi\ velocity field and integrated \hi\ spectrum of IC 5270 and NGC~7418. See captions of Figs. \ref{fig:cutouts0} to \ref{fig:cutouts2} for details. For IC~5270 we show the spectrum of the galaxy without the two \hi\ clouds in grey.}
\label{fig:ic5270}
\end{figure*}

The most interesting result of these observations is the detection of a few \hi\ clouds in the proximity of IC~5270 and NGC~7418 (Figs. \ref{fig:mosaic} and \ref{fig:ic5270}). In the former we detect two clouds. The brightest of the two is $\sim4$ arcmin north of IC~5270 and contains $1.6\times10^9$ \msun\ of \hi. The second cloud is $\sim7$ arcmin north-east of IC~5270 and its \hi\ mass is $1.0\times10^9$ \msun. The clouds account for $\sim1/3$ of the total \hi\ mass associated with IC~5270 in the ASKAP data (Table \ref{tab:memb}).

Available images at ultraviolet (GALEX), optical (DSS) and infrared (WISE) wavelengths do not reveal any bright counterparts to these \hi\ clouds. There are a few, very faint GALEX catalogued sources within $\sim1$ arcmin of both clouds, but their association with the \hi\ is not obvious. The main limitation is that the total integration time of the combined far- and near-UV GALEX image is just a few hundred seconds. A much deeper  observation is available for the group central region but IC~5270 is just outside this field. Similarly, DSS optical imaging is not very sensitive, and the deep optical image made by \cite{malin1985} stops just south of IC~5270. Therefore, while we find no obvious stellar counterpart, deeper images may be necessary to definitively establish that the \hi\ clouds are not gas-rich dwarf neighbours of IC~5270.

Another possibility is that the clouds are truly ``dark'', and are the densest clumps of a larger, low-column-density \hi\ structure -- possibly made of \hi\ stripped from IC~5270. Star-less clouds with a similar \hi\ mass are known to exist around other galaxies, and our detections would not be exceptional in this respect \citep[e.g.,][]{kilborn2006,oosterloo2007}. These clouds are often found in the presence of other signs of tidal interaction, which we may not be able to see in existing, shallow images of IC~5270 (e.g., \citealt{koribalski2003,english2010,serra2013,lee2014}; for a simulations perspective see, e.g., \citealt{bekki2005a}). Some indications in favour of this hypothesis come from a joint analysis of ASKAP and HIPASS data.

Firstly, with the knowledge of our ASKAP results, the detection of the clouds becomes clear also in the HIPASS data. The HIPASS detection associated with IC~5270 is spatially resolved and extended towards the north-east. An elliptical Gaussian fit to the \hi\ image of this source returns a central position which is offset from the optical position of IC~5270 by 2.6 arcmin. This offset is significantly larger than for all other HIPASS sources in the field, which match the optical position of their host galaxy to better than 1 arcmin (in agreement with the distribution of offsets between optical and HIPASS positions presented for larger samples by \citealt{koribalski2004} and \citealt{wong2009}). Furthermore, the HIPASS spectrum is in good agreement with the sum of the ASKAP spectra of IC~5270 and the two clouds, as shown in Fig. \ref{fig:ic5270}. In that figure we also show the ASKAP spectrum of IC~5270 alone (grey line) -- something that we can only do thanks to the higher angular resolution of these data. The obvious conclusion is that the \hi\ clouds are responsible for the strong peak above 1900 \kms, which gives the spectrum its unusually asymmetric appearance. While the Parkes data alone were not sufficient to unambiguously resolve the clouds from \hi\ in IC~5270 on the sky or in velocity, the new ASKAP data allow us to interpret unambiguously the shape of the HIPASS spectrum.

\begin{figure*}
\includegraphics[width=18cm]{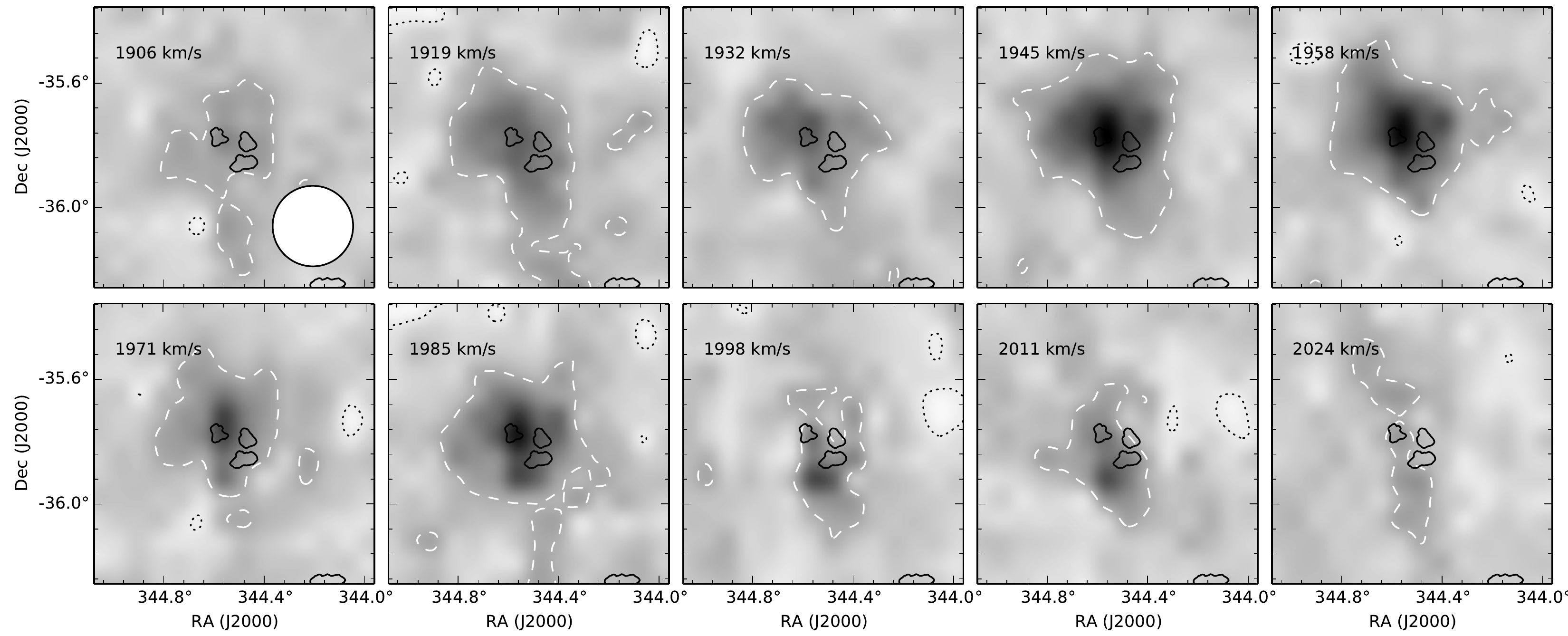}
\caption{Channel maps of the residual cube obtained subtracting the ASKAP \hi\ cube from the HIPASS \hi\ cube in the region around IC~5270. The ASKAP cube was convolved with the 15.5 arcmin HIPASS beam (represented by the white circle in the top-left panel) and regridded to the coordinate grid of the HIPASS cube in order to produce the residual cube. The solid black contour represents the ASKAP \hi\ image and is identical to the red contour in Fig. \ref{fig:mosaic}. Black-dotted and white-dashed contours represent residual emission at a level of $-30$ and $+30$ \mJybeam, respectively. The recessional velocity of the channels is indicated in the top-left corner.}
\label{fig:hipassres}
\end{figure*}

Further insight comes from the observation that the HIPASS spectrum contains some additional \hi\ emission missing from the ASKAP spectrum at velocities between 1900 and 2000 \kms\ (Fig. \ref{fig:ic5270}). What is the nature of this emission, and does it tell us anything important about this system? We investigate this aspect by convolving and regridding the ASKAP cube to the resolution and coordinate grid of the HIPASS cube, and subtracting the former from the latter\footnote{A possible caveat in this procedure is that we assume the HIPASS PSF to be Gaussian. In fact, the PSF is not exactly Gaussian, and its shape depends on the S/N and shape of the \hi\ source \citep{barnes2001}.}. We show channel maps of the residual cube in Fig. \ref{fig:hipassres}. Despite the low angular resolution, the figure reveals that the excess HIPASS emission is clearly located outside the main body of IC~5270. This emission must have a column density below $\sim10^{20}$ cm$^{-2}$, else it would have been detected in the ASKAP cube. We conclude that the \hi\ clouds detected by ASKAP are just the densest clumps of a diffuse, extended distribution of \hi, which according to the Parkes spectrum should contain an additional $\sim10^9$ \msun\ of gas.

We note that at least part of this faint, diffuse \hi\ emission is detected by the shortest baseline in our observation (37 m), and is visible in the natural-weighted cube made including this baseline. However, cleaning this emission proves extremely challenging because of the high, broad PSF sidelobes (see Sec. \ref{sec:hiimaging}). The full ASKAP will sample the inner regions of the $uv$ plane in a much more complete way and will therefore be able to image such emission with relative ease.

The other galaxy where we detect an \hi\ cloud is NGC~7418. The \hi\ disc of this object has similar size as the stellar disc, and we detect the cloud at the disc's north-west edge (Fig. \ref{fig:ic5270}). The cloud is clearly a separate gas system as it is kinematically distinct from the disc rotation. It hosts $\sim6\times10^8$ \msun\ of \hi, about 13 percent of the total, and its velocity is close to systemic. As in the case of IC~5270, we find no obvious stellar counterpart to the \hi\ cloud in available GALEX, DSS and WISE images. However, in this case the GALEX image has a relatively long exposure ($\sim1700$ s), and no stellar counterpart is visible in the deep optical image by \cite{malin1985} either. This suggests that the cloud is not a dwarf companion of NGC~7418.

Similar to the case of IC~5270, the Parkes spectrum of NGC~7418 shows some excess emission relative to the ASKAP spectrum (see velocities between 1400 and 1500 \kms\ in Fig. \ref{fig:ic5270}), and the HIPASS detection is resolved by the 15.5 arcmin PSF. Therefore, it is possible that the \hi\ cloud of NGC~7418, too, is embedded in an underlying, low-column-density gas distribution. Observations with better \hi\ column-density sensitivity would be needed to explore this possibility in more detail and to investigate the connection between this cloud and the diffuse intra-group \hi\ detected further north by \cite{walsh1990} and \cite{oosterloo1999}.

\section{Summary and conclusions}
\label{sec:summary}

We present \hi\ imaging of the IC~1459 galaxy group obtained with six ASKAP antennas equipped with phased-array feeds. We detect \hi\ in 11 galaxies down to a column density of $\sim10^{20}$ cm$^{-2}$ within a field of $\sim6$ deg$^2$, with a resolution of $\sim1$ arcmin on the sky and $\sim8$ \kms\ in velocity. These are the first resolved \hi\ images for 6 of the 11 detections.

Our data reveal the presence of a $\sim6\times10^8$ \msun\ \hi\ cloud in the proximity of NGC~7418 and of two $\sim10^9$ \msun\ clouds around IC~5270. The former amounts to $\sim13$ percent of the total \hi\ associated with NGC~7418. The two clouds around IC~5270 make up $\sim1/3$ of the total \hi\ mass associated with this galaxy by low resolution \hi\ observations, and explain the asymmetric Parkes \hi\ spectrum of this object. Based on a comparison between ASKAP and HIPASS data we conclude that the clouds around IC~5270 are the densest clumps of a larger \hi\ distribution below the column density sensitivity of the ASKAP data. We estimate this diffuse gas system to contain $\sim10^9$ \msun\ of gas in addition to the mass of the two clouds. The cloud around NGC~7418, too, may be part of a larger distribution of low-column-density gas.

The detection of intra-group \hi\ nearby IC~5270 as well as NGC~7418 adds to the body of evidence suggesting significant interaction between galaxies and their environment in the IC~1459 galaxy group. Previous results include the detection of intra-group, low-column-density \hi\ between the early-type IC~1459 and the peculiar NGC~7418A \citep{walsh1990,oosterloo1999} and evidence of ram-pressure stripping in NGC~7421 \citep{ryder1997}. Altogether, the total mass of \hi\ residing outside galaxies in this group is several times $10^9$ \msun. This is at least 10 percent of the \hi\ mass contained inside galaxies in the group. This estimate should be seen as a conservative lower limit because of the limited column density sensitivity of our data. Given that we are observing just a short snapshot in the long assembly history of this particular galaxy group, our estimate implies a substantial flow of \hi\ in and out of group galaxies over a period of several billion years.

Overall, these results demonstrate the good performance of phased-array feeds and give a taste of what the full ASKAP will be able to do on much larger areas of the sky. Considering its larger number of antennas, lower $T_\mathrm{sys}/\eta$ and larger number of simultaneous beams, the full ASKAP will be able to reach the same column-density sensitivity of our 30-h observation (inclusive of the shortest baseline) over a 5 times larger area of the sky and with twice the angular resolution in just a few hours. Longer integrations over large areas will make it be possible to measure the mass of \hi\ outside galaxies -- and firmly establish or rule out evidence of gas accretion and stripping -- for thousands of galaxies like the ones studied here.

\section*{Acknowledgments}

We acknowledge useful discussions with Phil Edwards and Ron Ekers. The Australian SKA Pathfinder is part of the Australia Telescope National Facility  which is funded by the Commonwealth of Australia for operation as a National Facility managed by CSIRO. This scientific work uses data obtained from the Murchison Radio-astronomy Observatory (MRO), which is jointly funded by the Commonwealth Government of Australia and State Government of Western Australia. The MRO is managed by the CSIRO, who also provide operational support to ASKAP. We acknowledge the Wajarri Yamatji people as the traditional owners of the Observatory site. This work was supported by iVEC through the use of advanced computing resources located at The Pawsey Centre. Parts of this research were conducted by the Australian Research Council Centre of Excellence for All-sky Astrophysics (CAASTRO), through project number CE110001020.

\bibliographystyle{mn2e}
\bibliography{../myrefs}

\label{lastpage}

\end{document}